\documentclass[journal]{IEEEtran}
\usepackage[utf8]{inputenc}
\usepackage{booktabs}
\usepackage{lscape} 
\usepackage{placeins}
\usepackage{amsmath}
\usepackage{graphicx}
\usepackage{pifont}
\usepackage{amssymb}
\usepackage{tabularx}
\usepackage{tabto}
\usepackage{graphicx}
\usepackage{multirow}
\usepackage{balance}

\graphicspath{ {./images/} }

\title{False Data Injection Attacks in Smart Grids:\\ State of the Art and Way Forward}

\author{
    \IEEEauthorblockN{
        Muhammad Irfan\IEEEauthorrefmark{1}\textsuperscript{\textsection}, 
        Alireza Sadighian\IEEEauthorrefmark{1}\textsuperscript{\textsection}, 
        Adeen Tanveer\IEEEauthorrefmark{1}, 
        Shaikha J. Al-Naimi\IEEEauthorrefmark{1},
        Gabriele Oligeri\IEEEauthorrefmark{1}}\\
    \IEEEauthorblockA{
        \IEEEauthorrefmark{1} 
            Division of Information and Computing Technology\\
            College of Science and Engineering, 
            Hamad Bin Khalifa University\\
            Qatar Foundation, Doha, Qatar.\\
    \emph{\{muir45306,asadighian,adta33802,shal45176,goligeri\}@hbku.edu.qa}
    }
}
\date{March 2023}

\begin{document}
\maketitle

\begingroup\renewcommand\thefootnote{\textsection}
\footnotetext{Both authors contributed equally to this work.}
\endgroup

\begin{abstract}
%In the recent years, cyber-attacks to smart grids are becoming more frequent, while proving their detrimental effect against the cyber and physical layers of the smart grid. Among the many malicious activities that can be launched against a smart grid, False Data Injection (FDI) attacks have raised significant concerns from both academia and industry as they are considered as a major threat to the internal state of the smart grid. FDI attacks can affect the (internal) state estimation process---critical for smart grid monitoring and control---thus being able to bypass conventional Bad Data Detection (BDD) methods. Hence, prompt detection and precise localization of FDI attacks is becomming of paramount importance to ensure smart grids security and safety. Several papers recently started to study and analyze this topic from different perspectives and address existing challenges. Data-driven techniques and mathematical modeling are the major ingredients of the proposed approaches. 
In the recent years, cyber-attacks to smart grids
are becoming more frequent.
Among the many malicious activities that can be launched against
smart grids, False Data Injection (FDI) attacks have raised
significant concerns from both academia and industry. FDI attacks can affect the (internal) state estimation
process—critical for smart grid monitoring and control—thus
being able to bypass conventional Bad Data Detection (BDD)
methods. Hence, prompt detection and precise localization of FDI
attacks is becomming of paramount importance to ensure smart
grids security and safety. Several papers recently started to study
and analyze this topic from different perspectives and address
existing challenges. Data-driven techniques and mathematical
modelings are the major ingredients of the proposed approaches.

The primary objective of this work is to provide a systematic review and insights into FDI attacks joint detection and localization approaches considering that other surveys mainly concentrated on the detection aspects without detailed coverage of localization aspects. For this purpose, we select and inspect more than forty
major research contributions, while conducting a detailed analysis of their methodology and objectives in relation to the FDI attacks detection and localization. We provide our key findings of the identified papers according to different criteria such as employed FDI attacks localization techniques, utilized evaluation scenarios, investigated FDI attack types, application scenarios, adopted methodologies and the use of additional data. Finally, we discuss open issues and future research directions. 
\end{abstract}


\section{Introduction}
\label{sec:introduction}

Over the last years, power systems are experiencing the migration to smart grids, in particular with the increasing popularity of Industrial Internet-of-Things (IIoT), which has moved to its next stage. However, this infrastructural upgrade has introduced a number of major challenges. Vulnerability to cyber-attacks is perhaps the most important challenge among all others. Adversaries exploit such vulnerabilities to manipulate system parameters and compromise the underlying infrastructure resulting into various physical and economical damages~\cite{yu2016smart}. Hence, real-time monitoring of smart grids is essential to ensure their healthy states. This involves analyzing continuous data-stream from various measurement devices deployed throughout the system, which are topologically distributed and structurally interrelated.

Among the various cyber-attacks to smart-grids, False Data Injection (FDI) raises greater concerns as advised by National Institute of Standards and Technology (NIST)~\cite{pillitteri2014guidelines}. Indeed, unlike other attack types, the underlying smart grid appears to operate as usual since the FDI attack affects the smart grid state estimation by injecting false data. FDI attacks can be deployed to the different layers of the smart grid, such as physical layer, communication layer, network layer and process layer~\cite{musleh2019survey}. Recently emerged FDI attacks are considered as a major challenge to smart grids state estimation as they can bypass conventional Bad Data Detection (BDD) methods~\cite{deng2016false}. 

Defending against FDI attacks has received a lot of attention by researchers. A few surveys have been devoted to detection solutions based on data-driven algorithms and mathematical modelings. However, these works do not provide information about FDI attacks localization. Indeed, beside detection, identifying the location of the attack is critical in order to define and issue effective countermeasures, and also, performing preventive actions. Therefore, recent research has focused on the joint task of detection and localization of FDI attacks. However, a major challenge that those solutions encounter is related to their performance, i.e., most attack localization methods have intrinsic difficulties locating the specific region that is affected by the FDI attack, in particular considering real-time constraints and accuracy requirements. Indeed, proposed solutions have high computational complexity and are difficult to apply to large-sized smart grids.

This survey presents a detailed review of recent research on FDI attack localization in smart grids with particular emphasis to recent challenges, enhancements and extension methods. To this aim, we identified a comprehensive list of papers that address this research challenge in various ways. Next, we analyzed this list from different perspectives, such as adopted techniques, application scenario, type of their input data, evaluation techniques, type of their underlying power systems, etc., and present our key findings and insights. 

%todo: after Survey comparison and results
%{\bf Purpose and scope of the study.} FDI attack is widely studies in the are of power systems. A well structured FDIA can bypass the existing BDD techniques. A series of process is conducted to detect any malicious data in fed into the system is not, first conduct the SE and then feed the meter measurements along with the SE to BDD mechanism.  

{\bf Contributions.} The contributions of this paper are manifold. Firstly, we identified state of the art research papers related to smart grids FDI detection and localization. We summarize each research paper describing its methodology, evaluation technique and future directions. Subsequently, we identify two main categories of contributions: data-driven approaches and mathematical modelings.
We describe our findings and insights according to the different categories, such as application scenarios, input types, adopted techniques, and input datasets. Finally, we present the limitations and the existing challenges in the current state of the art while providing some future directions. 

%\dinge{52} for checkmark \ding{55} for crossmark %
% \usepackage{multirow}
% \usepackage{graphicx}
%\begin{table*}[]
%\caption{Comparison of Our Paper with the Existing Survey Papers}
%\label{tab:comparison-other-surveys}

\begin{table*}[]
\caption{Comparison with the existing surveys}
\label{tab:comparison-other-surveys}
\centering
\resizebox{\textwidth}{!}{%
\begin{tabular}{ccccccccccccc}
\hline
\multirow{2}{*}{\textbf{Author}} & \multirow{2}{*}{\textbf{Location Detection}} & \multirow{2}{*}{\textbf{FDIA Detection}} & \multicolumn{5}{c}{\textbf{Methods}} & \multirow{2}{*}{\textbf{Detection Rate/Accuracy}} & \multirow{2}{*}{\textbf{Comparison with Other   Surveys}} & \multirow{2}{*}{\textbf{Comparative Study of   Algorithms}} & \multirow{2}{*}{\textbf{Attackers Knowledge   Levels}} & \multirow{2}{*}{\textbf{Covered Years}} \\ \cline{4-8}
 &  &  & \textbf{DL} & \textbf{GNN} & \textbf{GSP} & \textbf{MM} & \textbf{ML} &  &  &  &  &  \\ \hline
\multirow{3}{*}{Our   Survey} & \multirow{3}{*}{\ding{52}} & \multirow{3}{*}{\ding{52}} & \multirow{3}{*}{\ding{52}} & \multirow{3}{*}{\ding{52}} & \multirow{3}{*}{\ding{52}} & \multirow{3}{*}{\ding{52}} & \multirow{3}{*}{\ding{52}} & \multirow{3}{*}{\ding{52}} & \multirow{3}{*}{\ding{52}} & \multirow{3}{*}{\ding{52}} & Full & \multirow{3}{*}{2009-2023} \\
 &  &  &  &  &  &  &  &  &  &  & Partial &   \\ \hline
\multirow{2}{*}{Aoufi \emph{et. al.} \cite{aoufi2020survey}} & \multirow{2}{*}{\ding{55}} & \multirow{2}{*}{\ding{52}} & \multirow{2}{*}{\ding{52}} & \multirow{2}{*}{\ding{55}} & \multirow{2}{*}{\ding{55}} & \multirow{2}{*}{\ding{52}} & \multirow{2}{*}{\ding{52}} & \multirow{2}{*}{\ding{55}} & \multirow{2}{*}{\ding{52}} & \multirow{2}{*}{\ding{55}} & Full & \multirow{2}{*}{2009-2019} \\
 &  &  &  &  &  &  &  &  &  &  & Partial &  \\ \hline
Cui \emph{et. al.} \cite{cui2020detecting} & \ding{55} & \ding{52} & \ding{52} & \ding{55} & \ding{55} & \ding{55} & \ding{52} & \ding{55} & \ding{55} & \ding{55} & - & 2013-2019 \\ \hline
\multirow{3}{*}{Deng \emph{et. al.} \cite{deng2016false}} & \multirow{3}{*}{\ding{55}} & \multirow{3}{*}{\ding{52}} & \multirow{3}{*}{\ding{55}} & \multirow{3}{*}{\ding{55}} & \multirow{3}{*}{\ding{52}} & \multirow{3}{*}{\ding{52}} & \multirow{3}{*}{\ding{52}} & \multirow{3}{*}{\ding{55}} & \multirow{3}{*}{\ding{55}} & \multirow{3}{*}{\ding{55}} & Full & \multirow{3}{*}{2009-2014} \\
 &  &  &  &  &  &  &  &  &  &  & Partial &  \\
 &  &  &  &  &  &  &  &  &  &  & Zero   Knowledge &  \\ \hline
Guan \emph{et. al.} \cite{guan2015comprehensive}& \ding{55} & \ding{52} & \ding{55} & \ding{55} & \ding{52} & \ding{52} & \ding{52} & \ding{55} & \ding{55} & \ding{55} & Full & 2009-2013 \\ \hline
Husnoo \emph{et. al.} \cite{husnoo2022false} & \ding{55} & \ding{52} & \ding{52} & \ding{52} & \ding{52} & \ding{52} & \ding{52} & \ding{55} & \ding{52} & \ding{55} & - & 2013-2021 \\ \hline
Liang \emph{et. al.} \cite{liang2016review} & \ding{55} & \ding{52} & \ding{55} & \ding{55} & \ding{55} & \ding{52} & \ding{52} & \ding{55} & \ding{55} & \ding{55} & Full & 2009-2015 \\ \hline
\multirow{3}{*}{Musleh \emph{et. al.} \cite{musleh2019survey}} & \multirow{3}{*}{\ding{55}} & \multirow{3}{*}{\ding{52}} & \multirow{3}{*}{\ding{52}} & \multirow{3}{*}{\ding{55}} & \multirow{3}{*}{\ding{52}} & \multirow{3}{*}{\ding{52}} & \multirow{3}{*}{\ding{52}} & \multirow{3}{*}{\ding{52}} & \multirow{3}{*}{\ding{55}} & \multirow{3}{*}{\ding{52}} & Full & \multirow{3}{*}{2011-2019} \\
 &  &  &  &  &  &  &  &  &  &  & Partial &  \\
 &  &  &  &  &  &  &  &  &  &  & Zero   Knowledge &  \\ \hline
\multirow{3}{*}{Reda \emph{et. al.} \cite{reda2022comprehensive}} & \multirow{3}{*}{\ding{55}} & \multirow{3}{*}{\ding{52}} & \multirow{3}{*}{\ding{52}} & \multirow{3}{*}{\ding{55}} & \multirow{3}{*}{\ding{52}} & \multirow{3}{*}{\ding{52}} & \multirow{3}{*}{\ding{52}} & \multirow{3}{*}{\ding{55}} & \multirow{3}{*}{\ding{52}} & \multirow{3}{*}{\ding{55}} & Full & \multirow{3}{*}{2009-2021} \\
 &  &  &  &  &  &  &  &  &  &  & Partial &  \\
 &  &  &  &  &  &  &  &  &  &  & Zero   Knowledge &  \\ \hline
\multirow{3}{*}{Sayghe \emph{et. al.} \cite{sayghe2020survey}} & \multirow{3}{*}{\ding{55}} & \multirow{3}{*}{\ding{52}} & \multirow{3}{*}{\ding{52}} & \multirow{3}{*}{\ding{55}} & \multirow{3}{*}{\ding{55}} & \multirow{3}{*}{\ding{55}} & \multirow{3}{*}{\ding{52}} & \multirow{3}{*}{\ding{52}} & \multirow{3}{*}{\ding{55}} & \multirow{3}{*}{\ding{55}} & Full & \multirow{3}{*}{2010-2020} \\
 &  &  &  &  &  &  &  &  &  &  & Partial &  \\
 &  &  &  &  &  &  &  &  &  &  & Zero   Knowledge &  \\ \hline
\end{tabular}%
}
\end{table*}

\textbf{Roadmap}. The rest of this paper is organized as follows. Section~\ref{sec:comparison} compares this work with other related survey from the litterature, Section~\ref{sec:background} is dedicated to the background knowledge about smart grids and their existing challenges, Section~\ref{sec:methodlogy} describes our review methodology introducing the list of the selected papers and a preliminary break-down into use cases, Section~\ref{sec:data-driven} and Section~\ref{sec:math-modeling} summarize our findings related to the data-driven and the mathematical-modeling based approaches, respectively, Section~\ref{sec:analysis} summarizes the key-findings of our survey, while Section~\ref{sec:open-issues} highlights open issues and future research directions. Finally, Section~\ref{sec:conclusion} draws some concluding remarks.

\section{Comparison with existing surveys}
\label{sec:comparison}

In this section, we identify a few related surveys covering FDI attacks analysis. Next, we compare our work with them in terms of key contributions and extracted insights.

Reda \emph{et al.} in \cite{reda2022comprehensive} provide a comprehensive review of FDI attacks detection in smart grids. First, they provide a detailed description of different types of FDI attacks. Next, they analyze the impacts of these attacks on the target infrastructures. Moreover, this work studies the requirements for constructing FDI attacks, such as minimal attack vector, attack sparsity, incomplete parameters information, and requirements for unobservability of the system. The covered researches in this work mainly consider partial or full knowledge of the target system by the adversary.  

In \cite{husnoo2022false}, Husnoo \emph{et al.} study various FDI attacks targeting active power distribution systems. They propose a taxonomy of FDI attacks and classify them with respect to target components in smart grids. The authors also highlight shortcomings of the approaches proposed in selected studies, and propose recommendations to address them and mitigate corresponding threats. 

Aoufi \emph{et al.} in \cite{AOUFI2020} present a classification of different FDI attacks affecting State Estimation (SE), Security-Constrained Economic Dispatch (SCED), Security-Constrained Optimal Power Flow (SCOPF), Contingency Analysis (CA) in smart grids and discuss strategies to detect and mitigate each attack class. Moreover, this research highlights the challenges and open research issues related to FDI attacks including the need for more efficient and effective detection and mitigation techniques. 

Liang \emph{et al.} in \cite{liang2016review} study physical and economic impacts of FDI attacks and provide defense strategies. They also investigate the theoretical basis of FDI attack construction by highlighting four scenarios that an adversary can leverage to construct FDI attacks: (i) incomplete information, (ii) false topology, (iii) AC power flow model, and (iv) limited access to meters. 

Guan \emph{et al.} in \cite{guan2015comprehensive} classify FDI attacks based on Direct Current (DC) power model, Alternating Current (AC) power model and control system model. They provide countermeasures for each attack class. Moreover, they discuss existing problems in this field and provide future directions. 

An FDI attack can be considered as an anomaly in a smart grid to be detected using various data-driven (Artificial Intelligence (AI), Machine Learning (ML) and Deep Learning (DL)) techniques. Cui \emph{et al.} in \cite{cui2020detecting} review ML-based FDI attacks detection approaches and categorize them based on their target feature detection scenarios into state estimation, non-technical loss and load forecasting. Furthermore, they analyzes the issue of privacy leakage when using ML techniques for FDI attacks detection and mitigation, and propose potential countermeasures. 

Sayghe \emph{et al.} in \cite{sayghe2020survey} investigate data-driven approaches for detecting FDI attacks targeting state estimation algorithms. The main objective of this research is to overcome the limitations of traditional residual-based Bad Data Detection (BDD) approaches. 

Similarly but using different methods, Deng \emph{et al.} in \cite{deng2016false} investigate the detection of FDI attacks targeting state estimation. For this purpose, using mathematical and theoretical concepts, they analyze three main aspects of FDI attacks: (i) construction, (ii) defense, and (iii) impacts. They examine adversary perspectives in order to provide appropriate countermeasures. 

El-Nasser \emph{et al.} in \cite{youssef2018false} present a summary of research on FDI attacks against the state estimation in smart grids. They discuss theoretical principles of these attacks and classify these attacks based on the attacker knowledge amount and parameters of target power systems.

Table~\ref{tab:comparison-other-surveys} compares our work with existing surveys in the literature and highlights key differences and contributions. The main shortcoming of the above surveys is that they only concentrate on FDI attacks detection while not focusing on FDI attacks localization. To the best of our knowledge, this work is the first effort to conduct a comprehensive survey on FDI attacks joint detection and localization. Our work covers all the recently published data-driven and mathematical-modeling based approaches from deep learning and machine learning to graph signal processing (GSP) and graph neural networks (GNN). Moreover, our analysis compares the proposed solutions and provides insights from various dimensions and perspectives, such as attacker knowledge level, existing datasets, simulation environments, and future directions to address existing challenges.

\section{Background on Smart Grids}
\label{sec:background}

\begin{figure}
    \centering
    \includegraphics[width=0.9\linewidth]{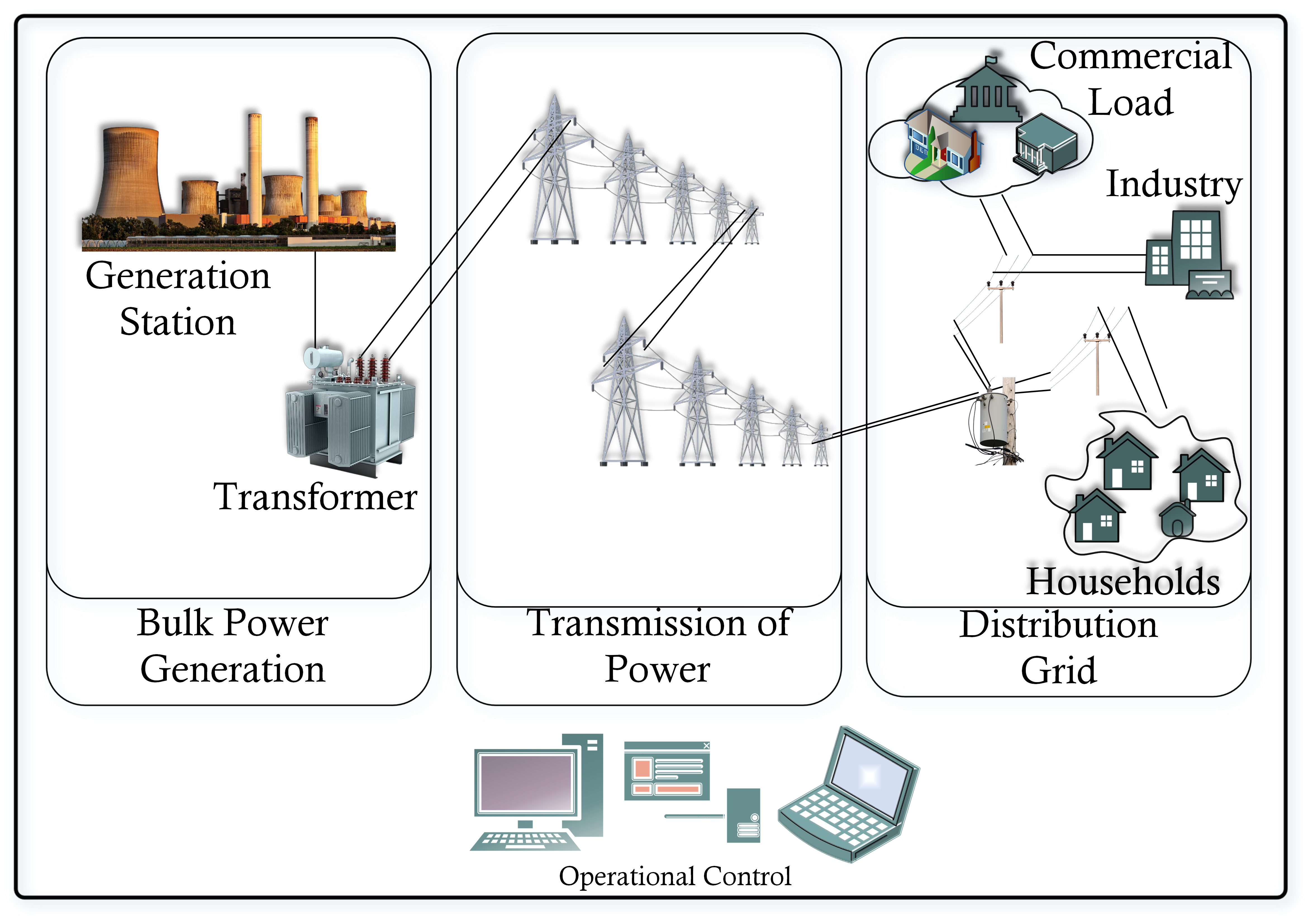}
    \caption{Architecture of a traditional electrical power grid.}
    \label{fig:traditional_power_grid}
\end{figure}

When considering a power system, the keyword grid is associated to an electrical system. A grid may perform some or all these four operations: (i) Energy Generation, (ii) Energy Transmission, (iii) Energy Distribution, and finaly, (iv) Energy control. The energy providers and consumers are connected in a network through transmission and distribution systems depending upon the operation. The architecture of a classical electrical power grid is shown in Fig.~\ref{fig:traditional_power_grid} where the electrical energy generated at the central power plant is delivered to end users by converting high voltage levels to lower voltage levels.

% Can we add some spaces on top of the labels Transmission Power (and the others)?

\begin{figure}
\centering
\includegraphics[width=\linewidth]{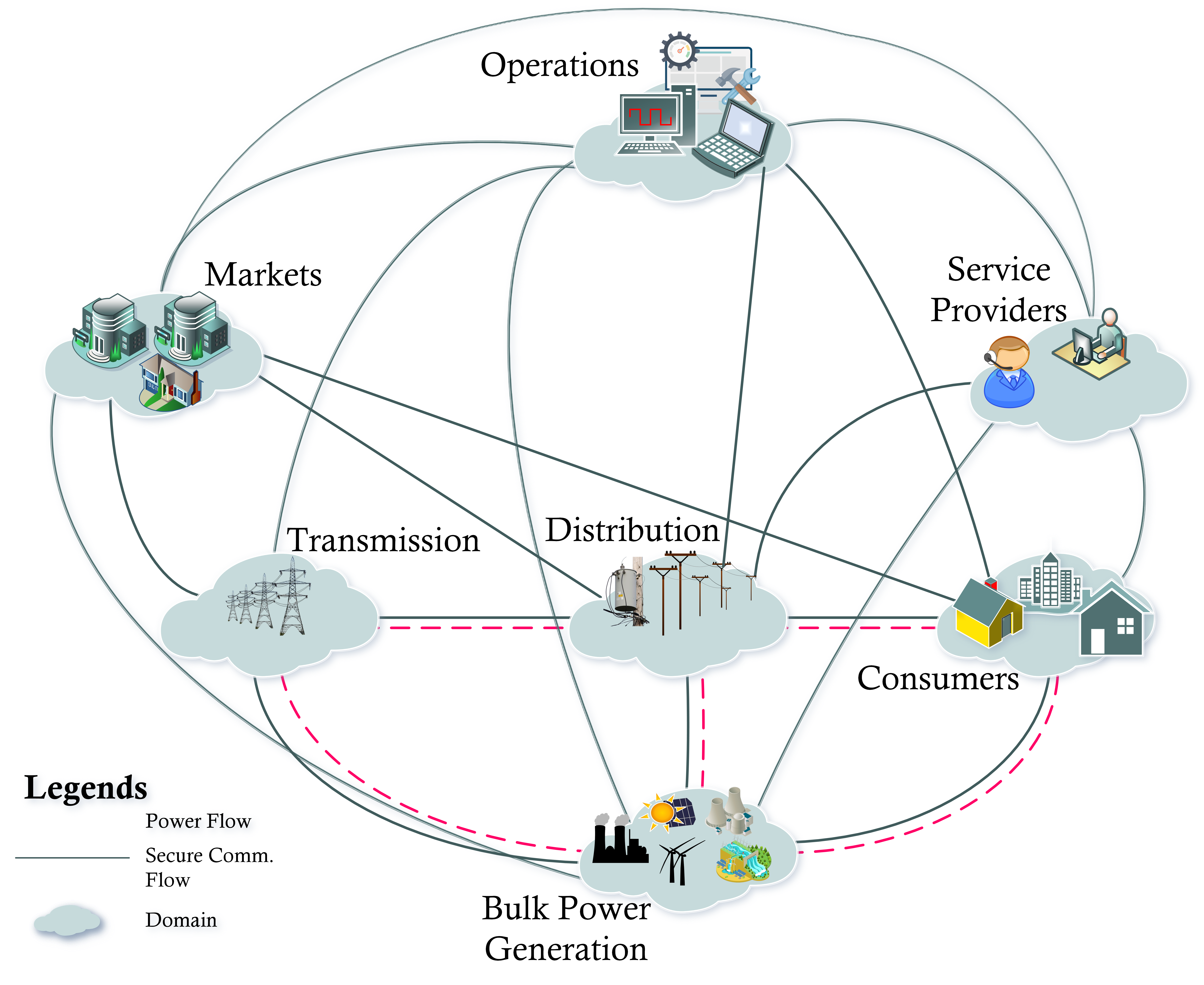}
\caption{Conceptual model of a smart grid.}
\label{fig:conceptual_model}
\end{figure}

A Smart Grid is an advanced version of a power grid where the devices are connected in order to share information among the domains defined by the National Institute of Standards and Technologies (NIST)~\cite{gopstein2021nist}. NIST conceptual model (Figure~\ref{fig:conceptual_model}) includes seven domains: (i) Customer, (ii) Markets, (iii) Generation including DER, (iv) Transmission, (v) Operation, (vi) Service Providers, (vii) Distribution. Table~\ref{tab:domain_roles_sg_conceptual_model} shows the type of service and roles of each domain. When we compare smart grids with traditional grids, significant improvements, such as type of communication, energy, monitoring, power generation, metering system, etc., can be highlighted. Table~\ref{fig:conceptual_model} presents a brief comparison between a traditional grid and a smart grid. 

\begin{table*}
\caption{Domains and roles/services in the smart grid conceptual model}
    \label{tab:domain_roles_sg_conceptual_model}
    \centering
    \begin{tabularx}{\textwidth}{|c|l|X|}
    \hline
         & \textbf{Domain} & \textbf{Services/Roles in the Domain} \\ \hline \hline
         1 & Customer & Three types of customer i- Residential, ii-Commercial, and iii- Industrial. These are the end users and may generate, manage and store the energy. \\ \hline 
         2 & Markets &   Intermediaries and participants in the power sector and other economic models that encourage activity and improve system results.                \\ \hline
         3 & Generation including DER &      The production and storage of electricity for later usage including Data Enery Resources (DER).   \\ \hline
         4 & Transmission &  The carrier of high power energy from one place to another over long distance.
         The may generate and store electricity. \\ \hline
         5 & Operations &   Electricity management  \\ \hline
         6 & Service Providers &  Provide services to end customer and utilities.              \\ \hline
         7 & Distribution &   The to and from distribution of electricity to end users (customers).                       \\ \hline
    \end{tabularx}
\end{table*}

\begin{table}
    \caption{Traditional grids versus Smart Grids}
    \centering
    \scriptsize
    \begin{tabular}{|l|l|l|}
    \hline
    & \textbf{Traditional Grid} & \textbf{Smart Grid} \\ \hline \hline
     \textbf{Communication} & One way    & Bi-Directional \\ \hline
     \textbf{Energy} & Electro-mechanical & Digital \\ \hline
     \textbf{Monitoring}   & Manual & Self\\ \hline
     \textbf{Power Generation} & Centralized & Distributed  \\ \hline
     \textbf{Control}   & Limited & Pervasive \\ \hline
     \textbf{Sensor}   & Limited & Throughout\\ \hline
     \textbf{Restoration}   & Manual & Self \\ \hline
     \textbf{Failure}   & Blackout & Adaptive and islanding\\ \hline
     \textbf{Structure} &  Hierarchical  & Network\\ \hline
     \textbf{Customers Involvement}   & No & Yes \\ \hline
     \multirow{2}{*} {\textbf{Metering System}}   & Electro-mechanical & Advanced Metering \\& & (Electronic)\\ \hline
    \end{tabular}
    
    \label{tab:trad_modern_sg}
\end{table}

\begin{figure}
    \centering
    \includegraphics[width=0.85\linewidth]{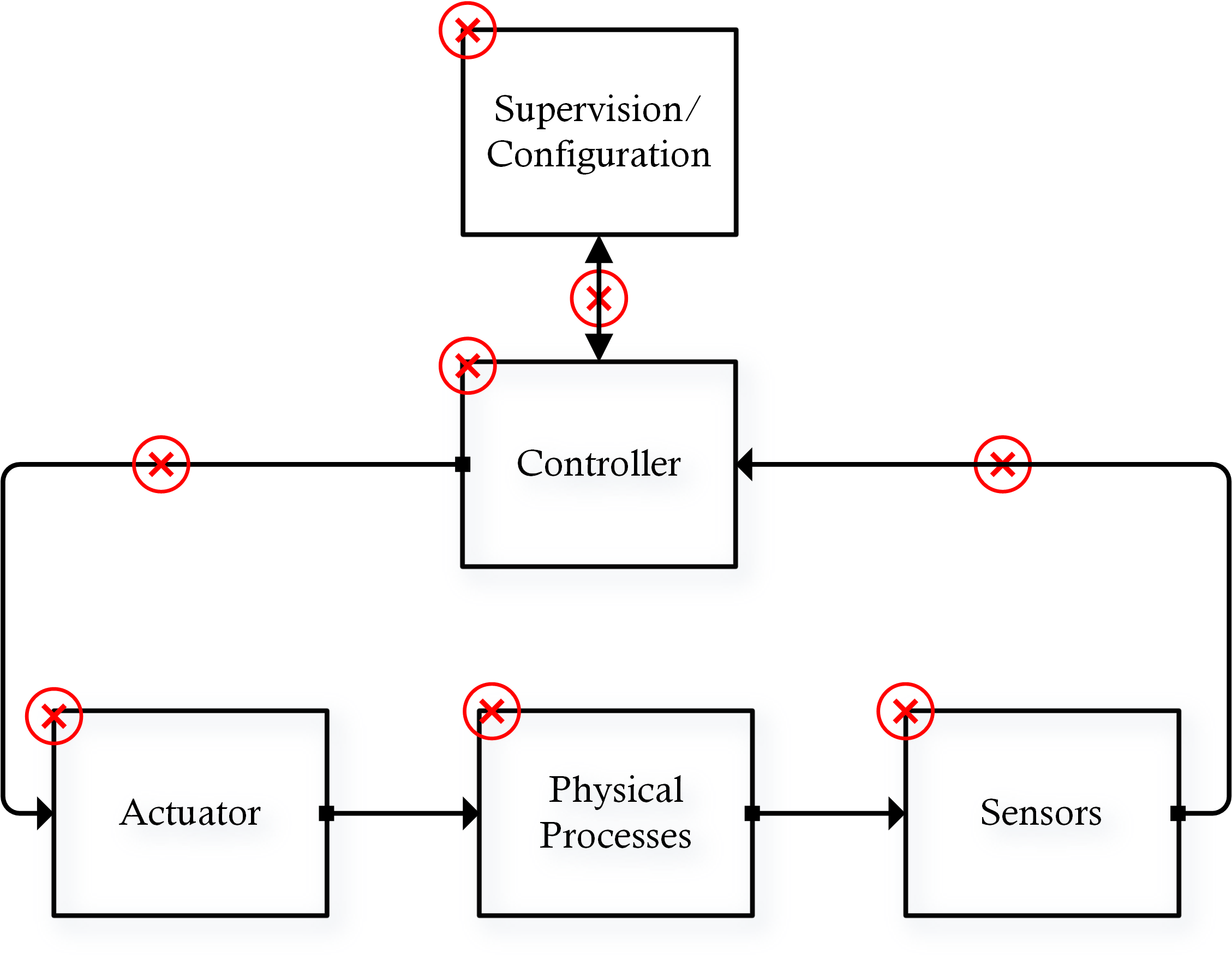}
    \caption{Possible attack sources in a smart grid \cite{cardenas2019cyber}.}
    \label{fig:entities_exposed_to_cyberthreats}
\end{figure}

Smart Grids provide real time access to users' data enabling them to control users' consumption patterns in an efficient way~\cite{Anuradha2013IEEEVision}. This exchange of information brings ease to better accommodate renewable energy sources (RES), and it improves the efficiency and reliability of their operations. However, such an inter-dependency communication technology makes smart grids susceptible to various types of vulnerabilities and cyber-attacks. In the recent years, several concerns were raised on the reliability and security of smart grids. The potential points which may be compromised are shown in Figure~\ref{fig:entities_exposed_to_cyberthreats}. To address these concerns, smart grids need to ensure essential security requirements, such as data integrity, availability, confidentiality and accountability, by all the key communication components.

\subsection{Smart Grid Components}
In this section, we provide a break-down of the smart grid key components.

\begin{itemize}
  \item[-] 
  {\em Remote terminal units} (RTU): an RTU is an electronic device that collects data periodically from the field-level devices and sends it to the control center, receives control commands from the control center, and transmits them to the field-level devices. RTUs can be targeted by various FDI attacks~\cite{konstantinou2016case}\cite{parizad2021laboratory}.
  %SCADA uses RTU's measurement for the reliable operation of the SG and research studies showed their susceptible nature towards FDI attacks \cite{konstantinou2016case}\cite{parizad2021laboratory}\cite{wang2021dual}.
  \item[-] {\em Phase measurement unit} (PMU): Phasor measurement units (PMUs) are devices that are used to provide synchronized measurements of voltages and currents in real-time. By sampling voltage and current waveforms simultaneously, synchronization is accomplished using timing signals from the Global Positioning System Satellite (GPS). These measurements provide information regarding the state of the network. PMU devices can be targeted by attackers to conduct different attacks, such as time synchronization and Data Manipulation attacks 
  %\cite{wang2019online}
  \cite{shereen2022detection}\cite{khalaf2018joint}.
  \item[-] {\em Power system state estimation} (PSSE): PSSE modules utilize information provided by RTUs and PMUs to perform current operating status estimations by statistical projections. FDI attacks on PSSEs are well studied in researches, such as \cite{ganjkhani2019novel}
  %\cite{liu2011false}
  \cite{an2022data}.
  \item[-] {\em Advanced metering infrastructure} (AMI): AMI is an integrated system of smart meters, communication networks, and data management systems that providing two-way communication between utilities and customers. AMI serves as an intermediary between consumers and utilities and is constituted of three tiers: (i) Home Area Network (HAN) inside the customer's home premises, (ii) Neighbor Area Network (NAN) connects multiple HANs, and (iii) Wide Area Network (WAN) connects NANs and HANs to utility center. Several attack types may target AMIs, such as Collusion attacks \cite{ibrahem2020pmbfe}\cite{alsharif2019epda}, FDI attacks \cite{na2021fake}, 
  %MiTM, 
  and replay attacks \cite{Saxena2017secure}.
  
  \item[-] {\em Advanced distribution management systems} (ADMS): ADMS is a software platform supporting the full suite of distribution management and optimization. It provides features that automate outage restoration and enhance the performance of the distribution grid.   
  \item[-] {\em Advanced transmission systems} (ATS): ATS are the most advanced systems and technologies to transfer electrical power. These systems seek to reduce losses and costs while improving power transmission efficiency, reliability, and sustainability. They address a variety of obstacles, such as the integration of renewable sources, the increase of capacity, and the improvement of power quality.
    \item[-] {\em High-voltage direct current} (HVDC): HVDC technology uses direct current (DC) for providing efficient means of transmitting large amount of power over long distances. It supports and improve the sustainability, efficiency and reliability of power supply systems. Amir \emph{et. al.} \cite{gholami2019cyber} performed analysis of the vulnerabilities, and impact of cyberattacks on the HVDC system.
    \item[-] {\em Advanced customer systems} (ACS): ACS is a set of digital platforms and tools power utilities use to engage with their customers and provide them with personalized information about their energy usage and costs. By facilitating more efficient and sustainable energy systems, advanced customer systems improve the customer experience.
\end{itemize}

\subsection{State Transition}

\begin{figure}
    \centering
    \includegraphics[width=0.8\linewidth]{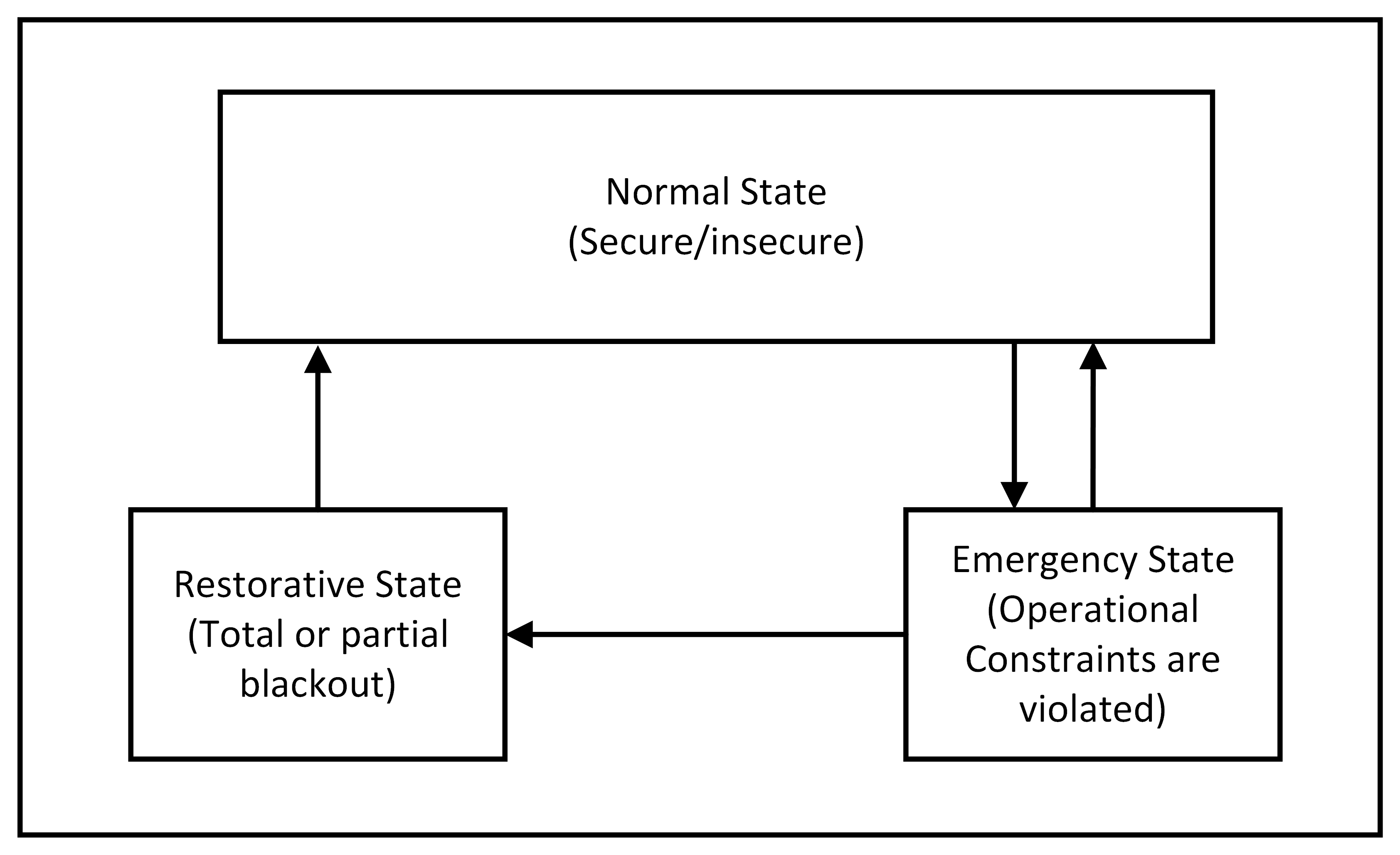}
    \caption{State transitions of a generic power systems}
    \label{fig:transition_states}
\end{figure}

A typical power system can be characterized by three internal states, i.e., Normal, Emergency, and Restorative. 

\begin{itemize}
  \item[-] \textbf{Normal}: a power system is considered normal when there is no violation of operational constraints, such as the limit of transmission line flows and upper/lower bounds of the magnitude of the voltages~\cite{abur2004power}. A list of critical contingencies is provided, and a normal state is treated as secure if it follows the occurrence of each contingency. These contingencies can be due to natural or unexpected failures to the transmission or generator outage. On the other hand, the normal state is known as insecure where all the operating inequality constraints and power balance at each bus are still satisfied. Preventive measurements are taken when the system is in normal but insecure in operating states.
\item[-] \textbf{Emergency}: the operating conditions of a power system may change and result in the violation of some operating constraints. At this stage, the power system still provides power to the loads. This Emergency state needs immediate attention to solve the underlying issue and bring the system back into its normal state. 
\item[-] \textbf{Restorative}: the preventive measurements taken by the emergency state may disconnect lines, loads, and other equipment to save the system from collapse. The system starts moving toward stability by removing the operating constraints. 
\end{itemize}

Figure~\ref{fig:transition_states} illustrates the 3 states and the associated transitions among them in a generic power system. Continuous monitoring of the system is required to ensure the reliable operations of the power grids. For this purpose, all the tasks associated with state estimation play a critical role in monitoring the reliability of operations of the underlying system. 

\subsection{False Data Injection Attacks}

A large percentage of smart grid cyber-attacks are categorized under FDI attacks. Compared to ordinary bad data attacks, FDI attacks are considered as data integrity attacks that inject false data misleading the PSSE measurements toward wrong states, and consequently, causing various systematic failures~\cite{yuan2011modeling}, such as key lines overloading and load shedding. More specifically, injected false data by an FDI attack disrupts the integrity of measurement data and forces PSSE to converge to a wrong operating point. This state miscalculation caused by injected data will mislead the smart grid operator and cause systematic problems. 

Once an FDI attack is detected in a smart grid, the tampered measurements should be recovered to ensure subsequent normal operations. Unfortunately, traditional Bad Data Detection (BDD) techniques are not able to detect and localize modern FDI attacks and they can be easily bypassed by recent FDI attacks~\cite{davis2012power}\cite{liu2011false}. Consequently, researchers continuously try to address such challenges by proposing modern and high-end FDI attack detection, mitigation, and localization approaches based on recently emerged technologies.

\section{Review Methodology}
\label{sec:methodlogy}

In this section, we describe our review methodology by firstly considering the list of papers taken into account for this survey, describing their approaches, and analytical details. In order to identify the list of considered papers, we used the following keywords: "Smart Grid", "False Data Injection", "Detection", "Identification" and "Localization". The major sources to collect the papers were: IEEE Xplore, ACM, SCOPUS, and Web of Science. 

\begin{figure}
    \centering
    \includegraphics[width=0.8\columnwidth]{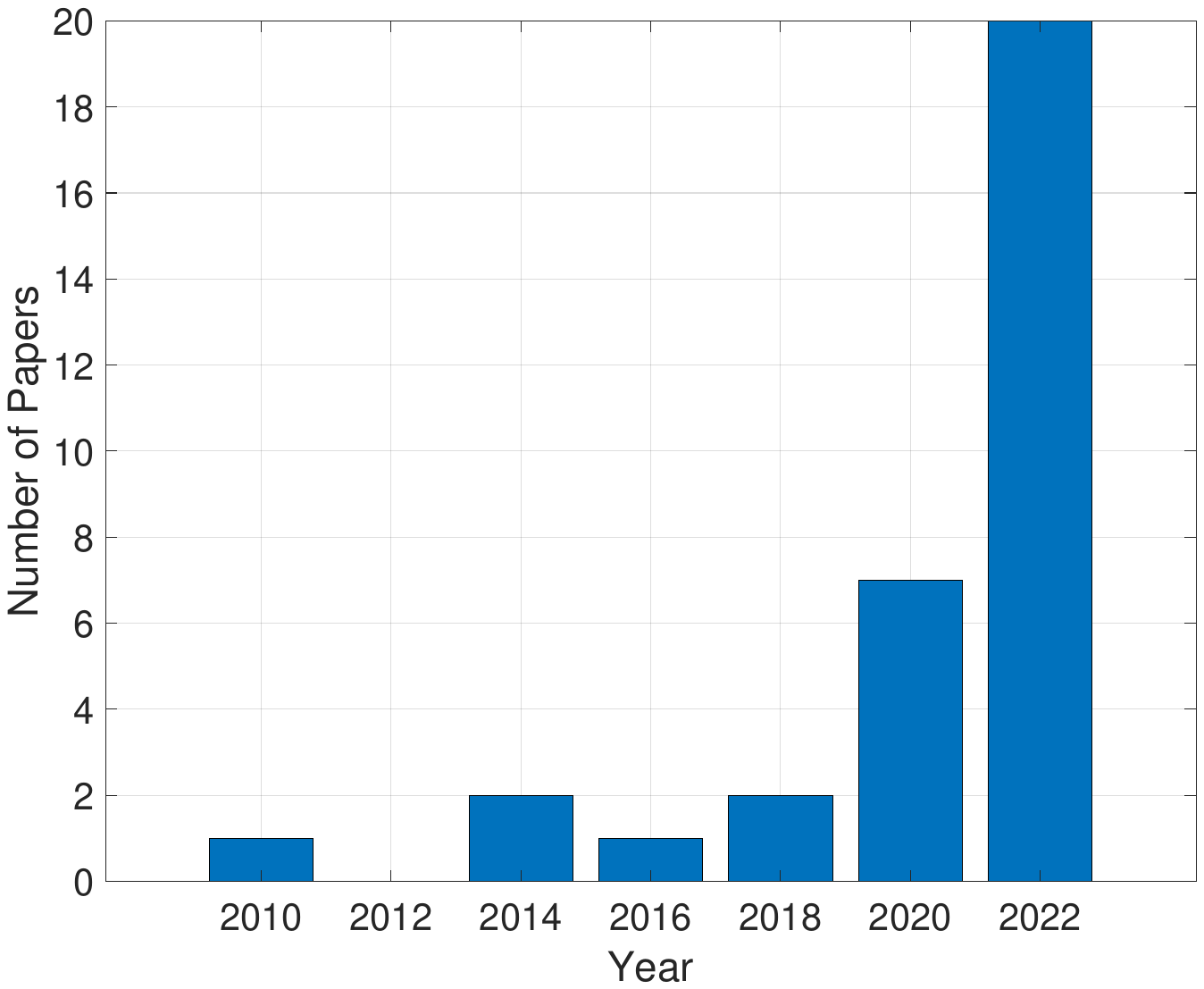}
    \caption{Number of papers related to FDI attacks} 
    \label{fig:paper-numbers-over-time}
\end{figure}

Table~\ref{tab:main-list-table} shows the list of the considered papers, where we report the title of the paper, the year of the publication, the authors' country, and finally, the number of citations collected up to the writing of this survey. Finally, in order to highlight the importance of the topic, Fig.~\ref{fig:paper-numbers-over-time} shows the number of FDI attack localization papers published between 2009 and 2022. There is a noticeable increase in the number of papers starting the year 2021 as part of the trend involving the wide application of machine learning and deep learning techniques to this field.

\begin{table*}[]
\caption{Our reference list of papers}
\centering
\begin{tabular}{clclc}
{\bf Reference}       & {\bf Title} & {\bf Year} & {\bf Country} & {\bf N. Citations} \\
\hline
\hline
%\cite{zonouz2012scpse}        & SCPSE: Security-Oriented Cyber-Physical State Estimation for Power Grid Critical Infrastructures & 2012 & US & 201 \\
%\cite{james2018online}        & Online False Data Injection Attack Detection With Wavelet Transform and Deep Neural Networks & 2018 & HK & 193 \\
%\cite{khalaf2018joint}        & Joint detection and mitigation of false data injection attacks in AGC systems & 2018 & CA & 76 \\
%\cite{gu2013bad}        & Bad Data Detection Method for Smart Grids based on Distributed State Estimation & 2013 & CN & 76 \\
\cite{wang2020locational}        & \begin{tabular}[c]{@{}l@{}}Locational detection of the false data injection attack in a smart grid: A multilabel classification\\ approach\end{tabular} & 2020 & CN & 55 \\
\cite{luo2018observer}        & Observer-based cyber attack detection and isolation in smart grids & 2018 & CN & 31 \\
\cite{wang2020detection}        & Detection and isolation of false data injection attacks in smart grids via nonlinear interval observer & 2019 & CN & 28 \\
\cite{boyaci2021joint}        & \begin{tabular}[c]{@{}l@{}}Joint detection and localization of stealth false data injection attacks in smart grids using graph\\ neural networks \end{tabular}& 2021 & US & 25 \\
\cite{nudell2015real}        & \begin{tabular}[c]{@{}l@{}}A Real-Time Attack Localization Algorithm for Large Power System Networks Using Graph-\\Theoretic Techniques \end{tabular}& 2015 & US & 24 \\
\cite{mukherjee2022deep}        & \begin{tabular}[c]{@{}l@{}}Deep learning-based multilabel classification for locational detection of false data injection attack \\in smart grids \end{tabular}& 2022 & IN & 20 \\
\cite{luo2020interval}        & Interval observer-based detection and localization against false data injection attack in smart grids & 2020 & CN & 20 \\
%\cite{jevtic2018physics}        & \begin{tabular}[c]{@{}l@{}}Physics- and learning-based detection and localization of false data injections in automatic \\generation control \end{tabular}& 2018 & US & 15 \\
\cite{siamak2020dynamic}        & \begin{tabular}[c]{@{}l@{}}Dynamic GPS Spoofing Attack Detection, Localization, and Measurement Correction Exploiting \\PMU and SCADA \end{tabular}& 2021 & IR & 14 \\
\cite{vukovic2013detection}        & Detection and localization of targeted attacks on fully distributed power system state estimation & 2013 & SE & 14 \\
\cite{luo2019detection}        & Detection and isolation of false data injection attack for smart grids via unknown input observers & 2019 & CN & 11 \\
\cite{smith2009event}        & Event detection and location in electric power systems using constrained optimization & 2009 & US & 11 \\
\cite{ganjkhani2021integrated}        & Integrated Cyber and Physical Anomaly Location and Classification in Power Distribution Systems & 2021 & US & 10 \\
\cite{hasnat2020detection} &  Detection and locating cyber and physical stresses in smart grids using graph signal processing    & 2020 & US & 10 \\
\cite{sakhnini2021physical}        & \begin{tabular}[c]{@{}l@{}}Physical layer attack identification and localization in cyber–physical grid: An ensemble deep\\ learning based approach \end{tabular}& 2021 & CA & 10 \\
\cite{shi2018pdl} &  PDL: An efficient prediction-based false data injection attack detection and location in smart grid   & 2018 & CN & 9 \\
\cite{mohammadpourfard2021attack} &  Attack Detection and Localization in Smart Grid with Image-based Deep Learning   & 2021 & TR & 6 \\
\cite{li2020deep} &  Deep Learning Based Covert Attack Identification for Industrial Control Systems   & 2021 & US & 6 \\
\cite{drayer2019cyber} &  Cyber Attack Localization in Smart Grids by Graph Modulation   & 2019 & IL & 4 \\
\cite{khalafi2021intrusion} &  Intrusion Detection, Measurement Correction, and Attack Localization of PMU Networks   & 2021 & IR & 4 \\
\cite{wang2013detection} &  Detection, recognition, and localization of multiple attacks through event unmixing   & 2013 & US & 4 \\
\cite{shereen2022detection} &  Detection and Localization of PMU Time Synchronization Attacks via Graph Signal Processing  & 2022 & SE & 3 \\
\cite{hasnat2021detecting} &  \begin{tabular}[c]{@{}l@{}}Detecting and locating cyber and physical stresses in smart grids using the k‐nearest neighbour\\ analysis of instantaneous correlation of states \end{tabular}  & 2021 & US & 2 \\
\cite{jiang2019location} &  Location of False Data Injection Attacks in Power System  & 2019 & CN & 2 \\
\cite{li2021locating} &  Locating False Data Injection Attacks on Smart Grids Using D-FACTS Devices  & 2021 & CN & 1 \\
\cite{hasnat2022graph} &  \begin{tabular}[c]{@{}l@{}}A Graph Signal Processing Framework for Detecting and Locating Cyber and Physical Stresses \\in Smart Grids  \end{tabular}& 2022 & US & 1 \\
\cite{qiu2021cyber} &  \begin{tabular}[c]{@{}l@{}}Cyber-attack localisation and tolerant control for microgrid energy management system based on\\ set-membership estimation \end{tabular} & 2021 & US & 1 \\
\cite{li2022adaptive} &  Adaptive Hierarchical Cyber Attack Detection and Localization in Active Distribution Systems  & 2022 & US & 1 \\
\cite{hasnat2021} &  \begin{tabular}[c]{@{}l@{}}Characterization and Classification of Cyber Attacks in Smart Grids using Local Smoothness of \\Graph Signals \end{tabular} & 2021 & US & 1 \\
\cite{boyaci2022infinite} &  Infinite Impulse Response Graph Neural Networks for Cyberattack Localization in Smart Grids  & 2023 & US & 1 \\
\cite{gao2022fast} &  \begin{tabular}[c]{@{}l@{}}Fast economic dispatch with false data injection attack in electricity-gas cyber–physical system: \\A data-driven approach  \end{tabular}& 2022 & CN & 0 \\
\cite{hegazy2022online} &  \begin{tabular}[c]{@{}l@{}}Online Location-based Detection of False Data Injection Attacks in Smart Grid Using Deep\\ Learning \end{tabular} & 2022 & EG & 0 \\
\cite{han2023false} &  \begin{tabular}[c]{@{}l@{}}False data injection attacks detection with modified temporal multi-graph convolutional network\\ in smart grids \end{tabular} & 2022 & CN & 0 \\
\cite{li2022online} &  An Online Approach to Covert Attack Detection and Identification in Power Systems  & 2022 & US & 0 \\
\cite{fanlocational} &  Locational Detection of Data Integrity Attacks with Multi-Gate Mixture-of-Experts in Smart Grid  & 2022 & CN & 0 \\
\cite{haghshenas2022temporal} &  A Temporal Graph Neural Network for Cyber Attack Detection and Localization in Smart Grids  & 2023 & US & 0 \\
%\cite{Hallaji2022Astream}        & \begin{tabular}[c]{@{}l@{}}A Stream Learning Approach for Real-Time Identification of False Data Injection Attacks in \\Cyber-Physical Power Systems \end{tabular}& 2022 & CA & 0 \\
\hline
\end{tabular}
\label{tab:main-list-table}
\end{table*}

%gab: this should be re-generated.

\begin{figure}
    \centering
    \includegraphics[width=\columnwidth]{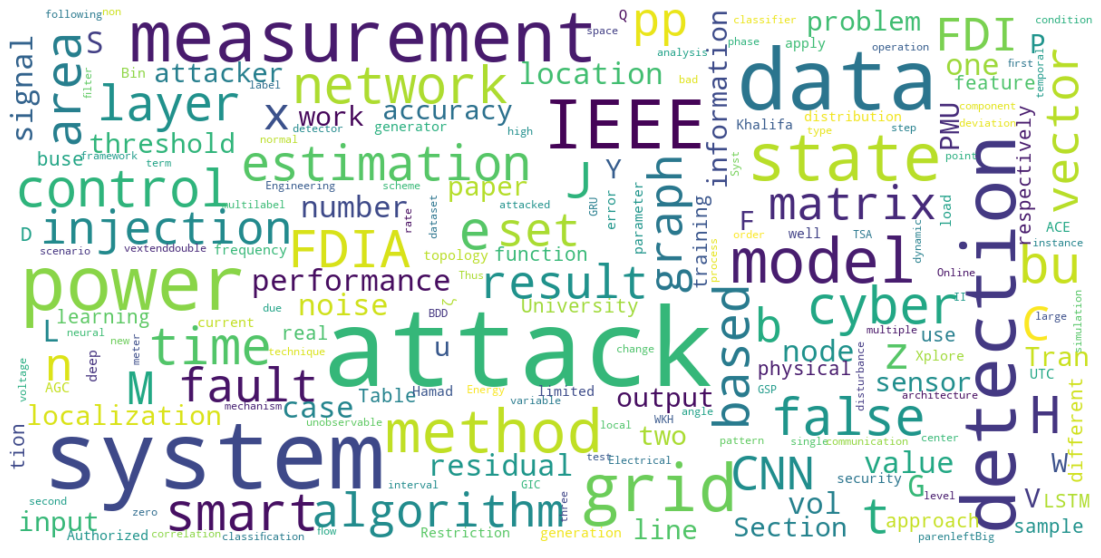}
    \caption{WordCloud computed over the set of papers considered for this survey} 
    \label{fig:wordcloud}
\end{figure}

Figure~\ref{fig:wordcloud} shows the word-cloud generated from the considered papers. While {\em Attack} is the most common word, as expected, other common words are {\em Detection}, {\em FDIA}, {\em State}, {\em Measurement}, {\em injection}, etc. which are core concepts in the framework, while concepts, such as {\em Graph}, {\em CNN}, {\em Matrix}, etc., are instantiated depending on the scenario application.

We initiate our analysis by categorizing the existing works into the following two general categories:

%gab: the classification below is not clear. Each benefit/limitation should be supported by a claim and the relative reference.

\begin{itemize}    
    \item[-] {\bf Mathematical modeling}. This class summarizes the methods based on mathematical and statistical estimators, e.g., Kalman filter~\cite{yong2016unified}), and aiming at the detection of known FDI attacks. 
    %Their performance degrade for unknown FDI attacks. 
    %Benefits of these methods are~\cite{musleh2019survey},~\cite{han2023false}: (i) no training required, (ii) no need for historical data, (iii) less memory is required compared to data-driven models. Main limitations of these methods are~\cite{musleh2019survey},~\cite{luo2020interval}: (i) need for system model and parameters, (ii) extensive computation, (iii) unscalable. 
    \item[-] {\bf Data-driven modeling.} These methods do not require prior knowledge and can detect unknown FDI attacks \cite{Hallaji2022Astream}. They are designed based on Artificial Intelligence (AI), Machine Learning (ML) and Deep Learning (DL) techniques. %Benefits of these approaches are~\cite{musleh2019survey},~\cite{Hallaji2022Astream}: (i) no need for system parameters, (ii) fast detection, (iii) real-time, (iv) scalable. Limitations of these methods are~\cite{musleh2019survey}: (i) extensive initial training, (ii) need for training dataset, (iii) extra memory space, (iv) over-fitting. 
\end{itemize}

\begin{figure}
    \centering
    \includegraphics[width=\linewidth]{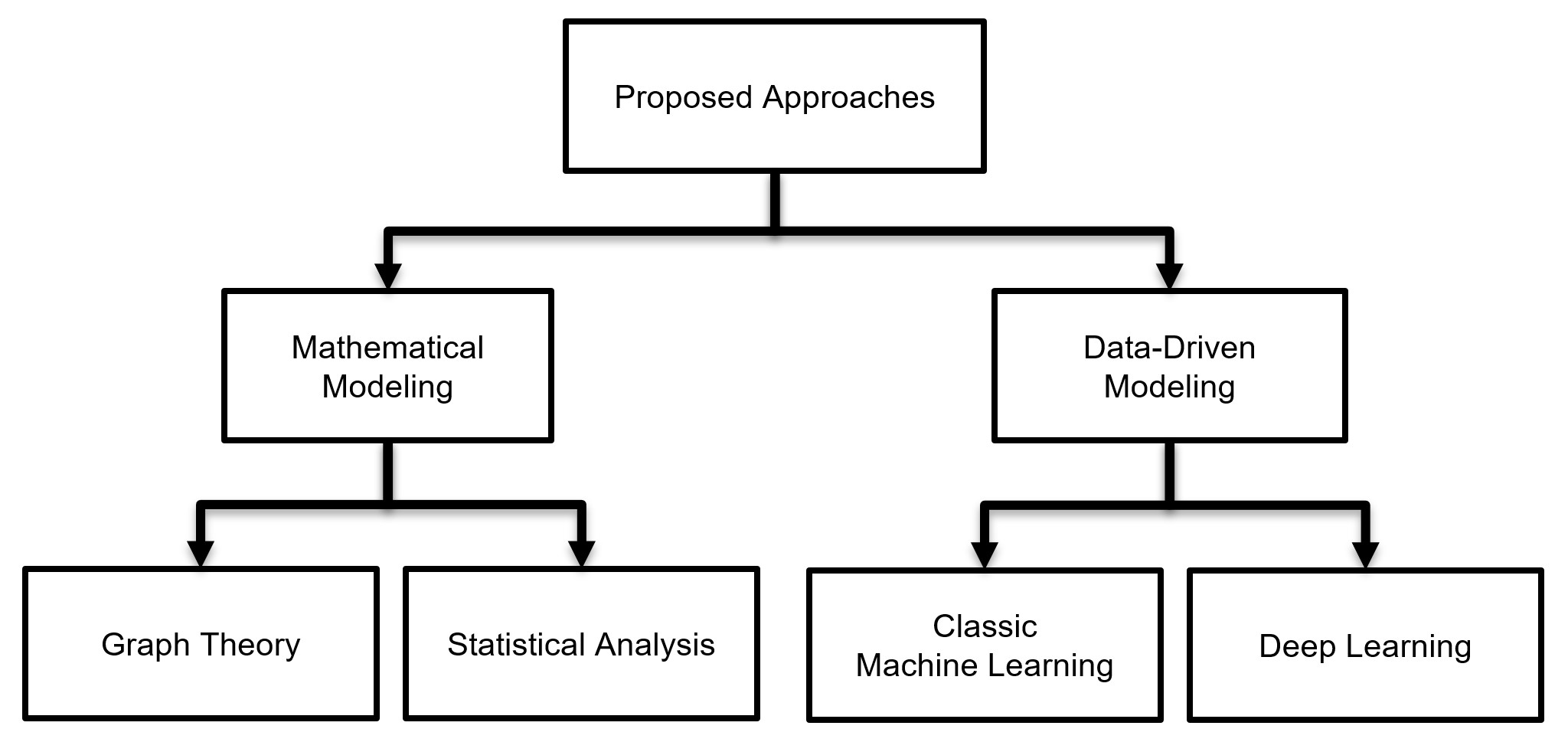}
    \caption{Categorization of the papers related to FDI attacks}
    \label{fig:categories}
\end{figure}

Figure~\ref{fig:categories} illustrates these categories along with their subcategories. Table~\ref{fig:use-case-table} also summarizes this categorization, while in the following, we map the papers to each identified category.

\begin{table}
\footnotesize
\centering
\caption{Use-case categories}
\begin{tabular}{|l|l|}
\hline
\textbf{Use Cases} & \textbf{Papers} \\ \hline\hline

\textbf{Data-Driven Modeling} & \begin{tabular}[c]{@{}l@{}}
\cite{hasnat2021detecting},
 \cite{jiang2019location},
 \cite{shereen2022detection},
 \cite{hasnat2021},
 \cite{li2022online},
 \cite{sakhnini2021physical},\\
 \cite{Hallaji2022Astream},
\cite{ganjkhani2021integrated},
 \cite{fanlocational},
 \cite{james2018online}
\cite{wang2020locational},
\cite{mohammadpourfard2021attack},\\
\cite{boyaci2021joint},
\cite{boyaci2022infinite},
\cite{haghshenas2022temporal},
\cite{han2023false},
\cite{li2020deep},
\cite{li2022adaptive},\\
\cite{gao2022fast},
\cite{jevtic2018physics},
\cite{hegazy2022online},
\cite{mukherjee2022deep}

\end{tabular}\\ \hline
 
\textbf{Mathematical Modeling} & \begin{tabular}[c]{@{}l@{}}
\cite{luo2018observer}, 
\cite{wang2020detection}, 
\cite{luo2020interval},
\cite{vukovic2013detection},
\cite{luo2019detection},
\cite{hasnat2020detection},\\
\cite{shi2018pdl},
\cite{drayer2019cyber},
\cite{qiu2021cyber},
\cite{wang2013detection},
\cite{hasnat2022graph},
\cite{li2021locating},\\
\cite{khalaf2018joint},
\cite{gu2013bad},
\cite{zonouz2012scpse}

\end{tabular}\\ \hline

\end{tabular}
\label{fig:use-case-table}
\end{table} 

\section{Data-Driven modeling}
\label{sec:data-driven}

In this section, we consider the papers addressing detection and localization of FDI attacks in smart grid using data-driven based modeling. Inspired by the recent advances in machine learning, and particularly, deep learning, these techniques have been widely employed by researchers to propose several solutions to identify and detect FDI attacks. Table~\ref{tab:data-driven} lists the data-driven models and the corresponding papers. 

\begin{table}
\footnotesize
    \centering
    \caption{Data-Driven Modeling}
    \begin{tabular}{|l|l|}
    \hline
        \textbf{Model/Technique} & \textbf{Papers} \\ \hline\hline

\textbf{Classic Machine Learning Techniques} & \begin{tabular}[c]{@{}l@{}}
\cite{hasnat2021detecting},
 \cite{jiang2019location},
 \cite{shereen2022detection},
 \cite{hasnat2021},\\
 \cite{li2022online},
 \cite{sakhnini2021physical},
 \cite{Hallaji2022Astream}
 \end{tabular}\\ \hline

\textbf{Deep Neural Networks} & \begin{tabular}[c]{@{}l@{}}
\cite{ganjkhani2021integrated},
 \cite{fanlocational},
 \cite{james2018online}
 \end{tabular}\\ \hline

\textbf{Convolutional Neural Networks} & \begin{tabular}[c]{@{}l@{}}
\cite{wang2020locational},
\cite{mohammadpourfard2021attack},
\end{tabular}\\ \hline

\textbf{Graph Neural Networks} & \begin{tabular}[c]{@{}l@{}}
\cite{boyaci2021joint},
\cite{boyaci2022infinite},
\cite{haghshenas2022temporal},
\cite{han2023false} 
\end{tabular}\\ \hline

\textbf{Recurrent Neural Networks} & \begin{tabular}[c]{@{}l@{}}
\cite{li2020deep},
\cite{li2022adaptive},
\cite{gao2022fast},
\cite{jevtic2018physics} 
\end{tabular}\\ \hline

\textbf{Combined Deep Learning Techniques} & \begin{tabular}[c]{@{}l@{}}
\cite{hegazy2022online},
\cite{mukherjee2022deep}
\end{tabular}\\ \hline

\end{tabular}
\label{tab:data-driven}
\end{table} 

Classic machine learning techniques are always the first data-driven approach to analyze the complexity of detecting and localizing cyber-attacks in smart grid environments. Hasnat \emph{et al.} in~\cite{hasnat2021detecting} utilize k‐nearest neighbour (kNN) classifiers to detect cyber-attacks in smart grids in a real-time mode. For this purpose, they detect and localize the anomalies by correlating patterns generated from power system buses leveraging correlation matrices. This matrix is an input to the KNN classifier for various cyber-attack detection. Jiang \emph{et al.} in~\cite{jiang2019location} propose a FDI attack detection and localization approach based on an ensemble tree based modeling, i.e., eXtreme Gradient Boosting (XGBoost) classifiers. In addition, they introduce a dimension extension model to record measurements corresponding locations. Shereen \emph{et al.} in~\cite{shereen2022detection} utilize Graph Signal Processing (GSP) and Random Forests (RF) classifiers that leverage GSP features to detect and localize Time Synchronization Attacks (TSAs) against Phasor Measurement Units (PMUs). Hasnat \emph{et al.} in~\cite{hasnat2021} study various cyber-attacks in smart grids and analyze them based on the local smoothness of their graph signals. Accordingly, they extract a group of local smoothness features to feed a neural network (NN) classifier. The results of the NN are used to characterise cyber-attacks and define corresponding profiles. Li \emph{et al.} in~\cite{li2022online} present a real-time covert attacks detection and localization approaches for smart grids based on the Sparse Group Lasso (SGL) algorithm. The underlying smart grid constitutes from multiple regional control centers (RCCs) vulnerable to covert attacks, and an independent system operator (ISO). The authors extend the proposed approach from linear to nonlinear system setups with relaxation. Sakhnini \emph{et al.} in~\cite{sakhnini2021physical} use an ensemble of machine learning techniques combining multiple classifiers (i.e. Decision Trees, Random Forest, feed-forward ANN) to propose an approach to detect and localize attacks in the smart grids' physical layer. In the proposed model, for attack localization, they use the chi-square metric to formulate attacks to specific features or measurements so that security experts can start mitigation process. Similarly, Hallaji \emph{et al.} in~\cite{Hallaji2022Astream} introduce a dynamic data-driven approach, called unobservable threat modelling using least-squares fitting (ORIGIN), helping power systems state estimations when they are under unobservable FDI attack. This approach relies on ensemble machine learning classifiers having 3 main phases: (i) FDI attacks detection based on observed and learned historical patterns, (ii) classification based extracted features to determine the type of attacks, (iii) retrieving control signals. 

Deep Neural Network (DNN) has also being considered by researchers for detecting and localizing FDI attacks. Ganjkhani \emph{et al.} in~\cite{ganjkhani2021integrated} used a fine-tuned DNN as a multi-class classifier to address the problem of localizing faults and cyber-attacks while being able to classify power system states and identify potential anomalies. The proposed solution (FALCON) adopts outage information issued by protection relays and fault indicators (FI) as DNN input features. Fan \emph{et al.} in~\cite{fanlocational} use DNNs to propose an FDI attack detection and localization approach formulated as a multi-task classification problem. For this purpose, they design a multi-gate mixture-of-experts (LD-MMoE) scheme considered as a multi-task classifier. The author believe that compared to the state-of-the-art approaches, this model is more computational-friendly as the number of trainable parameters are lower. A novel ac FDIA detection approach covering recent ac FDIA patterns is presented in~\cite{james2018online}. The proposed approach in addition to capturing spatial data features in the system state, it considers the temporal data correlation of the consecutive system states to identify adversarial patterns and activities. For this purpose, they build an intelligent FDIA detection system by employing discrete wavelet transform (DWT) algorithm for extracting the system state features in a given time slot and deep neural network (DNN) technique for analyzing the temporal-spatial characteristics of the data sequences (i.e. constructing a recurrent neural network, RNN).

Convolutional Neural Networks (CNN) have been widely used by researchers to propose various FDI attacks detection and localization approaches. Wang \emph{et al.} in~\cite{wang2020locational} introduce deep-learning-based localization and detection architecture (DLLD) that concatenates a CNN with a standard BDD used to filter low quality data. In the proposed architecture, the CNN model is a multi-label classifier aims to detect and locate abnormal behaviors and inconsistencies injected by FDI attacks. Inspired by recent advances in image processing, Mohammadpourfard \emph{et al.} in~\cite{mohammadpourfard2021attack} proposed a CNN based 2-stage framework for FDI attacks detection and localization. First, using Gramian Angular Field (GAF) and Recurrence Plot (RP), they encode temporal states of power systems into 2D images. Next, a fine-tuned CNN is used as a multi-label classifier to identify anomalies and corresponding locations. 

Considering the inherent graph structure of power systems that may not be modeled efficiently in the Euclidean space, graph neural networks (GNN) have been used in a number of power systems FDI attack detection and localization projects. In~\cite{boyaci2021joint}, Boyaci \emph{et al.} proposed a GNN based power grid FDI attack detection and localization approach that adopts the auto-regressive moving average (ARMA) type graph filters (GFs) to identify presence and location of inconsistencies. In~\cite{boyaci2022infinite}, Boyaci \emph{et al.} utilize Infinite Impulse Response (IIR) Graph Neural Networks (GNN) to detect and localize cyberattacks at the bus level. For this purpose, they compare IIR and FIR GFs and the results show that IIR GFs more accurately approximate the desired filter response. Haghshenas \emph{et al.} in~\cite{haghshenas2022temporal} introduce a Temporal GNN (TGNN) based framework for FDI attacks detection and localization in smart grids. In the proposed framework, the topological information of the underlying system are formulated into a GNN for abnormal measurements identification purpose, and temporal information at each bus of the underlying power system are formulated into a Gated Recurrent Unit (GRU). Based on their simulation results on the IEEE 118 bus system, they believe that the proposed framework is able to detect and localize FDI and ramp attacks with high accuracy. Han \emph{et al.} in~\cite{han2023false} present a FDI attack detection and localization method called modified temporal multi-graph convolutional network (MTMCN) that is a graph-theory based deep learning method. To feed this method, bus and branch related data are transformed into graph structured data representing topological data relationships. These information are correlated to build deep learning model's features and characteristics serving as a multi-label classifier. 

Recurrent neural networks (RNN) are used to capture the temporal correlations among system states. This capability of RNN has attracted researchers to use it for FDI attacks profiling. Li \emph{et al.} in~\cite{li2020deep} propose a hybrid data-driven approach that combines an autoencoder as a system status features extractor, a RNN to define system normal operation profile over time, and a DNN as a classifier to detect covert attacks presence and localize them. This hybrid framework by taking into account the spatial and temporal behavior of the underlying system, tries to reduce false alarm rate. In~\cite{li2022adaptive}, Li \emph{et al.} propose an adaptive hierarchical FDI attack detection and localization approach for power grid systems. The FDI attack detection task is performed by a Multi-layer Long Short-Term Memory Network (MLSTM). The FDI attack localization is a two-phase task that first, the FDI attack sub-regions are estimated, and next, to localize the FDI attack, the attack presence within the estimated sub-regions is calculated. Gao \emph{et al.} in~\cite{gao2022fast} present a data-driven approach incorporating a residual network (ResNet) and attention LSTM (ALSTM) in order to build a multi-label classifier for real-time FDI attacks locational detection (i.e. outlier locations of modified measurements by FDI attacks) in electricity-gas cyber–physical systems (EGCPS). The proposed system captures temporal data of measurements to detect and localize FDI attacks. 
Furthermore, they utilize a fast Dynamic Time Wrapping (DTW) for recovering tempered measurements. Jevtic \emph{et al.} in~\cite{jevtic2018physics} present a RNN based approach for detecting FDI attacks in Automatic Generation Controls (AGC). By analyzing historical data along with Area Control Error (ACE) data, an LSTM based neural network model is built and trained in order to predict ACEs. Any difference between the predicted ACEs and real ACE data reveals potential FDI attacks. 

Combining, concatenating or incorporating different data-driven models into a single FDI attack detection and localization system is another strategy utilized by researchers. Hegazy \emph{et al.} in~\cite{hegazy2022online}, first conduct a comparative and analytical study of various deep-learning based FDI attack localization approaches (e.g., CNN, LSTM, Bi-LSTM, CNN-LSTM) to represent their capabilities locating compromised measurements. Next, they propose a mixed model concatenating a multi-feature CNN (MCNN) and a LSTM models, called MCNN-LSTM, that analyzes multi-variate time-series measurements to detect and localize FDI attacks. Their results show that the proposed approach outperforms other deep learning models in term of resilience, scalability, complexity and prediction time. Similarly, Mukherjee1 \emph{et al.} in~\cite{mukherjee2022deep} conducted an analytical comparison among different machine learning and deep learning models, such as CNN, CNN-LSTM, CNN-GRU and KNN, for locational detection of structured and unstructured FDI attacks in smart grids. Based on their results, they believe that the conventional CNN model as an model-free multi-label classifier, outperforms other models in detecting inconsistencies caused by FDI attacks.

\section{Mathematical modeling}
\label{sec:math-modeling}

This section describes FDI attack detection and localization approaches based on mathematical modeling. Graph theory and statistical analysis are the two main sub-categories. In the following, we describe the research works of both sub-categories. Table~\ref{tab:math-modeling} lists the mathematical models and the corresponding papers.

A FDI attack detection and compensation approach based on simultaneous input and state estimation is proposed in~\cite{khalaf2018joint} for Automatic Generation Control (AGC) systems. In this approach, FDIA signal is considered as unknown input so that its value gets estimated for compensating the attack effect. Kalman filter~\cite{yong2016unified} is used for this purpose. The main benefits of this approach are: (i) it does not enforce any hardware upgrade or modification to the control center, (ii) no additional communication channel is needed compared to competitors. The authors to guarantee the decision accuracy of the AGC control center, for both the AGC system state and the adversarial signal calculate minimum variance unbiased estimate. Formal proofs are provided to ensure the stability of the AGC system after injecting the attack mitigation algorithm. Similarly, A FDI attack detection and isolation approach for large-scale smart grids is proposed in~\cite{wang2020detection}. The authors, first, define a group of non-linear interval observers trying to accurately estimate interval state of the system. Next, an interval residual-based detection metric to address the pre-computed threshold shortcomings is proposed. Once the interval residuals seems abnormal, an appropriate alarm is issued accordingly. Moreover, to isolate the sensors injecting FDIA, a matrix-based isolation algorithm is introduced and integrated to the control centers. In~\cite{luo2020interval}, Luo \emph{et al.} try to detect and localize FDI attacks based on a series of interval observers. FDI attack localization is performed by analyzing the measurement data of corresponding sensors located in different areas of the system. Hence, each interval observer is responsible for both detection and localization tasks. Smith \emph{et al.} in~\cite{smith2009event} explain that nonlinear constrained optimization (NLCO) can be used to estimate smart grids states from limited number of collected measurements. This can be used to detect and localize any anomaly (e.g., line outages, generator trips, load shedding, etc.) within the system. Wang \emph{et al.} in~\cite{wang2020detection} try to detect and isolate FDI attacks based on the unknown input (UI) interval observer. For this purpose, they design UI interval observers based on the physical dynamics of the underlying system. Next, they use the interval residuals of all areas to identify FDI attacks. For exceptional cases, when a FDI attack remains undetectable, a local matrix-based algorithm is developed to perform final judgments by combining observable cases. Shi \emph{et al.} in~\cite{shi2018pdl} propose prediction-based FDIA detection and location scheme (PDL) modeling power systems states as a multi-variate time-series predictable by vector autoregressive processes (VAR). Accordingly, a time-varying and non-diagonal matrix is prepared for analyzing states consistencies over time, and detecting potential anomalies (e.g., FDI attacks). After detecting and localizing anomalies, their values can be replace with corrected values for calibration purposes. Qiu \emph{et al.} in~\cite{qiu2021cyber} utilize set-membership estimation (SME) to detect and localize cyberattacks by calculating ellipsoidal estimates of the affected system states. Furthermore, they have designed a model predictive control (MPC) to compute optimal control sequences and improve system tolerance against cyberattacks. Wang \emph{et al.} in~\cite{wang2013detection} present "event unmixing" conceptual framework for online and temporal analysis (i.e. detection and localization) of cyberattacks in smart grids. By collecting temporal stamps to build a overcomplete dictionary, they characterise abnormal activities over time. Using this framework, they try to identify the type of cyberattacks, failure locations and failure start time. These information helps to efficiently localize and mitigate cyberattacks. 

In~\cite{luo2018observer}, Luo \emph{et al.} use real-time measurements to propose an state observer-based unknown attack detection and isolation approach. The main novelty of this work is integrating graph theory to large scale smart grids aiming to partition the system into a number of areas and identify sparse and dense ones. For this purpose, they design an observation matrix to extract the corresponding relation between the state observer residual and the smart grid inputs. Next, using the graph-theoretic algorithm, they calculate the asymmetric weighted Laplacian. Another graph-theoretic approach using real-time synchrophasor measurements is proposed in~\cite{nudell2015real} to localize cyber-attacks, such as Denial of Service (DoS), flooding, jamming of communication links, etc., in large-scale power systems. Leveraging physical topology of a smart grid, by dividing it into a set of areas or clusters, they use the phasor measurement data information to identify the area that a particular attack enters and isolate the attack. They compare the identified localization characteristics with a list of transfer function residues. Hasnat \emph{et al.} in~\cite{hasnat2020detection} utilize graph signal processing (GSP) as smart grids measurement data analysis and representation. They analyze cyberattacks (e.g., FDI, DoS, data-replay, etc.) impacts on smart grids in the graph vertex domain, the graph-frequency domain and combined vertex-frequency domain of signals. Next, based on the effects of attacks on the graph Fourier transform (GFT), they propose different GSP techniques for detecting and localizing attacks. similarly, Drayer \emph{et al.} in~\cite{drayer2019cyber} consider a power system as a graph where nodes represent buses and edges represent transmission lines. By analyzing smoothness of graph states, they try to detect and localize FDI attacks. 

\begin{table}
\footnotesize
\centering
\caption{MATHEMATICAL MODELING}
\begin{tabular}{|l|l|}
\hline
\textbf{Model/Technique} & \textbf{Papers} \\ \hline\hline

\textbf{Statistical Analysis} & \begin{tabular}[c]{@{}l@{}}
\cite{wang2020detection},
\cite{luo2020interval},
\cite{smith2009event},
 \cite{wang2020detection},
\cite{shi2018pdl},
\cite{yong2016unified}
  
\end{tabular}\\ \hline
 
\textbf{Graph Theory} & \begin{tabular}[c]{@{}l@{}}
\cite{luo2018observer}, 
\cite{nudell2015real}, 
\cite{hasnat2020detection},
\cite{drayer2019cyber}

\end{tabular}\\ \hline

\end{tabular}
\label{tab:math-modeling}
\end{table}

\section{Summary analysis and key findings}
\label{sec:analysis}

This section discusses the major findings of our analysis from different perspectives, such as the localization techniques, evaluation scenarios, adversary's knowledge level, result analysis, and finally, the investigated FDI attacks considered by the identified relevant works.

\subsection{Localization Techniques}

\begin{table}[]
\footnotesize
\centering
\caption{FDI Attacks Localization Techniques}
\label{tab:localization-technique}
%\resizebox{\columnwidth}{!}{%
\begin{tabular}{|l|l|}
\hline 
\textbf{Localization Technique} & \textbf{Papers} \\ \hline \hline

Classic Machine Learning& \begin{tabular}[c]{@{}l@{}}\cite{wang2020locational},~\cite{khalafi2021intrusion},~\cite{li2022adaptive},~\cite{hasnat2021detecting},~\cite{hasnat2021},~\cite{jiang2019location},~\cite{hasnat2021}\end{tabular} \\ \hline

Deep Learning& \begin{tabular}[c]{@{}l@{}}\cite{wang2020locational},~\cite{mukherjee2022deep},~\cite{ganjkhani2021integrated},~\cite{mohammadpourfard2021attack},~\cite{li2020deep},~\cite{han2023false},~\cite{hegazy2022online},\\\cite{fanlocational},~\cite{james2018online}\end{tabular} \\ \hline

%Neural Network & \begin{tabular}[c]{@{}l@{}}\cite{wang2020locational},~\cite{mukherjee2022deep},~\cite{ganjkhani2021integrated},~\cite{mohammadpourfard2021attack},~\cite{li2020deep},\\\cite{hegazy2022online},~\cite{han2023false},~\cite{fanlocational},~\cite{james2018online}\end{tabular} \\ \hline
Graph Neural Networks & \begin{tabular}[c]{@{}l@{}}\cite{boyaci2021joint},~\cite{boyaci2022infinite},~\cite{haghshenas2022temporal}\end{tabular} \\ \hline
%Clustering Based & \begin{tabular}[c]{@{}l@{}} \cite{khalafi2021intrusion},~\cite{li2022adaptive},~\cite{hasnat2021detecting},~\cite{hasnat2021}\end{tabular} \\ \hline
%Clustering Based & \begin{tabular}[c]{@{}l@{}}\cite{khalafi2021intrusion},~\cite{li2022adaptive},~\cite{hasnat2021detecting},~\cite{hasnat2021}\end{tabular} \\ \hline
Set-Membership Estimation & \begin{tabular}[c]{@{}l@{}}
\cite{qiu2021cyber}
\end{tabular} \\ \hline
Sparse Group Lasso & \cite{li2022online} \\ \hline
Chi-Square & \cite{sakhnini2021physical} \\ \hline
Graph Signal Processing &\cite{shereen2022detection},~\cite{hasnat2020detection},~\cite{hasnat2022graph} \\ \hline
Matrix Based & \begin{tabular}[c]{@{}l@{}}
\cite{wang2019detection},~\cite{luo2020interval},~\cite{shi2018pdl},~\cite{wang2013detection}\end{tabular} \\ \hline
Observer Based Residual & \begin{tabular}[c]{@{}l@{}}
\cite{luo2018observer},~\cite{luo2019detection}\end{tabular} \\ \hline

\end{tabular}%
%}
\end{table}

It is important to highlight the key differences/similarities found in the literature to locate an FDI attack or the region under the attack. In this section, we summarize the methodologies adopted by researchers to locate the node under an FDI attack.

Considering that the main focus of this survey is FDI attacks localization, we have listed the techniques utilized in the selected papers for localizing FDI attacks in Table~\ref{tab:localization-technique}. These techniques play an important role in particular when we want to control and mitigate FDI attacks. 
%Some extracted insights are the following:

% gab: each of the following claim should have a reference in one of the selected paper.

As Table~\ref{tab:localization-technique} shows, the majority of FDI attack localization researches employ advanced data-driven approaches revealing the direction and trend of this research area toward these approaches as also supported by Table~\ref{tab:main-list-table}. 
%the trend for the FDI attack localization research is toward using advanced data-driven approaches considering that the majority of recently published papers use this direction (Table~\ref{tab:main-list-table}). 
In this direction, the localization of the FDI attacks is mainly considered as a multi-label classification task or a multi-cluster clustering task. 

\subsection{Evaluation Scenario}

%gab: same as before, each claim should be supported by references

Various evaluation methodologies, such as statistical-based \cite{boyaci2021joint}\cite{wang2013detection}, simulation based \cite{drayer2019cyber}\cite{fanlocational}, emulation based \cite{sakhnini2021physical}, real test-bed based \cite{haghshenas2022temporal}\cite{Hallaji2022Astream} etc., have been considered to evaluate the proposed methodologies in the reviewed papers. 
%Each methodology has particular pros and cons. For example, statistical evaluation does not provide the best performance despite no initial training is required. Simulation based evaluation is highly adaptive and scalable. However, its quality directly depends on the quality of the simulation software. Real test-beds have higher trustiness but less flexibility, while resulting expensive. 

%In the following, we provide our findings and earned insights from those evaluation methodologies. 

A comprehensive dataset can facilitate and accelerate the evaluation process. Such data can either be taken directly from smart grids or derived synthetically. As of infrastructure privacy and confidentiality, very limited ground truth datasets are publicly available, e.g., New York Independent System Operator (NYISO)~\cite{tierney2010new}. Consequently, in most cases, the data is generated via simulations using various tools and platforms. Table~\ref{tab:datasets} shows the list of existing datasets and tools for data generation.  

\begin{table}
\footnotesize
\centering
\caption{Adopted datasets}
\begin{tabular}{|l|l|}
\hline
\textbf{Input} & \textbf{Papers} \\ \hline \hline
\textbf{Data Generated by MATPOWER} &  \begin{tabular}[c]{@{}l@{}} 
\cite{wang2020locational}, 
\cite{boyaci2022infinite}, 
\cite{mukherjee2022deep}, 
\cite{hasnat2020detection}, 

\\
\cite{shi2018pdl},
\cite{mohammadpourfard2021attack}, 
\cite{li2020deep}, 
\cite{hasnat2022graph}, 

\\
\cite{hasnat2021detecting}, 
\cite{jiang2019location}, 
\cite{shereen2022detection},
\cite{boyaci2022infinite},
\\
\cite{hasnat2021}, 
\cite{Lee2021}, 
\cite{haghshenas2022temporal}, 
\cite{gao2022fast}, 
\\
\cite{hegazy2022online}, 
\cite{han2023false},
\cite{li2022online}, 
\cite{fanlocational}, 
\\
\cite{gu2013bad},
\cite{zonouz2012scpse} \end{tabular}\\ \hline
\textbf{Data Generated by MATLAB} &  
\cite{luo2020interval},
\cite{ganjkhani2021integrated}\\ \hline
\textbf{Data from DigSilent Power Factory}\footnote{https://www.digsilent.de/en/powerfactory.html} &  
\cite{james2018online} \\ \hline
\textbf{NYISO Load Profile}\footnote{https://www.nyiso.com/load-data} &  
\cite{boyaci2022infinite},
\cite{hasnat2021detecting}, 
\cite{hasnat2021},
\cite{jevtic2018physics} \\ \hline
\textbf{Simulated frequency disturbance data} &  
\cite{wang2013detection} \\ \hline
\textbf{Simulated Waveform Data of IEEE-37 bus} &
\cite{li2022adaptive} \\ \hline
\begin{tabular}[c]{@{}l@{}} \textbf{Oak Ridge National}\footnote{https://www.ornl.gov/} \\ \textbf{Laboratories simulated dataset}\end{tabular} &
\begin{tabular}[c]{@{}l@{}} 
\cite{sakhnini2021physical} \end{tabular}\\ \hline
\textbf{PMU data} & 
\cite{Hallaji2022Astream} \\ \hline
%\textbf{Other} &  \begin{tabular}[c]{@{}l@{}}
%\cite{luo2018observer},
%\cite{wang2020detection},
%\cite{vukovic2013detection},
%\cite{luo2019detection},
%\cite{drayer2019cyber},
%\\
%\cite{khalafi2021intrusion},
%\cite{qiu2021cyber},
%\cite{khalaf2018joint}
%\end{tabular} \\ \hline
\end{tabular}
\label{tab:datasets}
\end{table}

From the simulation point of view, Table~\ref{tab:simulators} reports the most popular simulators considered in the literature, i.e., MATPOWER\footnote{https://matpower.org}, MATLAB simulink\footnote{https://www.mathworks.com/products/simulink.html}, Power System Simulator for Engineering (PSS/E)\footnote{https://www.siemens.com/global/en/products/energy/grid-software/planning/pss-software/pss-e.html}, MATLAB YALMIP~\cite{Lofberg2004}, or custom simulations. Moreover, the power system components from which the data is collected play an important role. The data can be read from different sources, such as a Phase Measurement Unit (PMU), a smart meter, a Frequency Disturbance Recorder (FDR), digital relays, etc. There is actually no best-practice since, in general, the sampling rate of a PMU is greater than that of a smart meter, i.e. data is collected with higher resolution, but reduced processing speed. Additionally, a smart meter can record basic amplitude whereas PMU gives phase information too. 

We now considered two reference scenarios, which are the result of the analysis of the paper considering the aforementioned tools.

\begin{enumerate}
\item \textbf{IEEE case studies:} most of the studies utilize standard IEEE-X system models for their experiments and evaluating their approaches. X represents the number of buses indicating the overall complexity of the power system. Table~\ref{tab:ieee-list-description} compares such standards in terms of number of lines, number of generators and number of busses while Table~\ref{tab:evaluation-scenario} matches the standards to the corresponding papers. IEEE-14 and IEEE-118 are the two most popular standards utilized by researchers.

%In the selected literature, we found that there exist no publically available real dataset, due to privacy issue for the evaluation of proposed studies.  Most of studies considered standard IEEE-X (X represents number of buses). In literature we found that a number of papers are considering IEEE-14 \cite{wang2020locational} \cite{mukherjee2022deep} \cite{shi2018pdl} \cite{drayer2019cyber} \cite{khalafi2021intrusion} \cite{li2020deep}\cite{jiang2019location} \cite{shereen2022detection} \cite{li2021locating} \cite{hegazy2022online} \cite{han2023false} \cite{li2022online} \cite{fanlocational}\cite{gu2013bad} \cite{boyaci2022infinite}. The selected proposed studies consider one or more IEEE-X bus systems to verify the generalization of their solutions. Following IEEE-14 bus system, IEEE-118 bus system is the second highest bus system considered by the researchers in selected literature \cite{james2018online}\cite{wang2020locational}\cite{boyaci2021joint}\cite{mukherjee2022deep}\cite{vukovic2013detection}\cite{mohammadpourfard2021attack}\cite{hasnat2022graph}\cite{hasnat2021detecting}\cite{boyaci2022infinite}\cite{hasnat2021}\cite{haghshenas2022temporal}\cite{hegazy2022online}\cite{han2023false}\cite{fanlocational}. In some studies large IEEE-300 distribution bus system is considered under the proposed solution such as \cite{james2018online}\cite{boyaci2021joint}\cite{boyaci2022infinite}. Some research studies considered IEEE-39\cite{shereen2022detection}, and IEEE-57\cite{boyaci2021joint}\cite{mohammadpourfard2021attack}. 

\item \textbf{Other case studies:}
beside the IEEE-X bus systems, there have been custom test cases examined by Luo \emph{et al.} in~\cite{luo2018observer} employed Two-Area Kundur System as a baseline with two areas and extended it to 180-generators with 10-areas. 

%In~\cite{qiu2021cyber}, they used MATLAB YALMIP~\cite{Lofberg2004} as a simulation tool to solve the set-membership estimation optimization problem. Wang \emph{et al.} in~\cite{wang2013detection} evaluated their proposed technique to detect and localize multiple attack locations using "Power System Simulator for Engineering" PSS/E \cite{siemensKnowPSSE}. This simulation has 23 buses, 6 generators, 7 loads and 17 branch lines.

\end{enumerate}

%Beside the IEEE-X bus systems, there has been custom test cases examined by the researcher. Author Luo et. al. \cite{luo2018observer} used Two-Area Kundur System as baseline with two areas. They extended the proposed solution to 180-generators with 10-areas. Author in \cite{qiu2021cyber} used MATLAB YALMIP\cite{Lofberg2004} as a simulation tool which solves the set-membership estimation optimization problem. Wang et. al.\cite{wang2013detection} evaluated their technique for detection and localization of multiple attack locations with "Power System Simulator for Engineering" PSS/E \cite{siemensKnowPSSE}. It is simulated model with 23 buses, 6 generators, 7 loads and 17 branch lines. In \cite{li2022adaptive} used IEEE 37-node model built with OPAL-RT, there is one voltage source inverter(VSI) based distributed energy source and two phtovotaic are connected main power grid.

\begin{table*}[]
\caption{IEEE Standard Bus System Comparison}
\centering
\begin{tabular}{cccccc}
 {\bf IEEE-X Bus System} & {\bf No. of Lines/Branches} & {\bf No. of Generators}  & {\bf No. of Slack Buses} & {\bf No. of Generator Buses} & {\bf No. of Load Buses} \\
\hline
  14 & 20 & 5 & 1 & 4 & 9 \\
  30 & 41 & 6 & 1 & 5 & 24 \\
  39 & 46 & 10 & 1 & 9 & 29 \\
  57 & 80 & 7 & 1 & 6 & 50 \\
  118 & 186 & 54 & 1 & 53 & 64 \\
  300 & 411 & 69 & 1 & 68 & 231 \\
  
\hline
\end{tabular}
\label{tab:ieee-list-description}
\end{table*}

\begin{table}
\footnotesize
\centering
\caption{Evaluation Scenario}
\begin{tabular}{|l|l|}
\hline
\textbf{Power System} & \textbf{Papers} \\ \hline \hline
        
\textbf{IEEE-14} &  \begin{tabular}[c]{@{}l@{}} 
\cite{wang2020locational},
\cite{mukherjee2022deep},
\cite{shi2018pdl},
\cite{drayer2019cyber},
\cite{khalafi2021intrusion},
\cite{li2020deep},
\cite{jiang2019location},
\cite{shereen2022detection},\\
\cite{li2021locating},
\cite{hegazy2022online},
\cite{han2023false},
\cite{li2022online},
\cite{fanlocational},
\cite{gu2013bad}  \end{tabular}\\ \hline

\textbf{IEEE-37} &  
\cite{li2022adaptive} \\ \hline

\textbf{IEEE-39} &  
\cite{shereen2022detection} \\ \hline

\textbf{IEEE-57} &  
\cite{boyaci2021joint},
\cite{mohammadpourfard2021attack} \\ \hline

\textbf{IEEE-118} &  \begin{tabular}[c]{@{}l@{}} 
\cite{james2018online},
\cite{wang2020locational},
\cite{boyaci2021joint},
\cite{mukherjee2022deep},
\cite{vukovic2013detection},
\cite{mohammadpourfard2021attack},
\cite{hasnat2022graph},
\cite{hasnat2021detecting},\\
\cite{boyaci2022infinite},
\cite{hasnat2021},
\cite{haghshenas2022temporal},
\cite{hegazy2022online},
\cite{han2023false},
\cite{fanlocational}   \end{tabular}\\ \hline

\textbf{IEEE-143} & 
\cite{ganjkhani2021integrated} \\ \hline
 
\textbf{IEEE-300} &  \begin{tabular}[c]{@{}l@{}} 
\cite{james2018online},
\cite{boyaci2021joint},
\cite{boyaci2022infinite}
\end{tabular}\\ \hline

\textbf{Two-Area Kundur} & 
\cite{luo2018observer} \\ \hline

\end{tabular}
\label{tab:evaluation-scenario}
\end{table}

\begin{table}
\footnotesize
\centering
\caption{Simulators}
\begin{tabular}{|l|l|}
\hline
\textbf{Environment} & \textbf{Papers} \\ \hline \hline
        
\textbf{MATPOWER} &  \begin{tabular}[c]{@{}l@{}} 
\cite{wang2020locational},
\cite{mukherjee2022deep},
\cite{shi2018pdl},
\cite{drayer2019cyber},
\cite{khalafi2021intrusion},
\cite{li2020deep},
\cite{jiang2019location},
\cite{shereen2022detection},\\
\cite{li2021locating},
\cite{hegazy2022online},
\cite{han2023false},
\cite{li2022online},
\cite{fanlocational},
\cite{gu2013bad},
\cite{boyaci2022infinite}  \end{tabular}\\ \hline

\textbf{MATLAB-Simulink} &
\cite{li2022adaptive} \\ \hline

\textbf{PSS/E} &
\cite{shereen2022detection} \\ \hline

\textbf{MATLAB YALMIP}\footnote{https://yalmip.github.io/} & 
\cite{qiu2021cyber} \\ \hline

\end{tabular}
\label{tab:simulators}
\end{table}

\subsection{Adversary knowledge}
%gab: even here, claims should be supported by references
%irf: Section 3. https://ietresearch.onlinelibrary.wiley.com/doi/epdf/10.1049/iet-stg.2020.0015 
% section

Adversary knowledge is constituted by a set of information that can be exploited to increase the effectiveness of the attack. Such knowledge can be any information about the smart grid infrastructure, components, and operations. Adversary knowledge can be categorized into two levels \cite{sayghe2020survey}: (i) full knowledge about the underlying infrastructure of the system, (ii) limited knowledge about the underlying infrastructure of the system. It is hard for an attacker to access full knowledge regarding the target smart grid system~\cite{li2020deep}~\cite{mukherjee2022deep}. These limitations are the result of (i) power subsystems being geographically far from each other, restricting the attacker from collecting information for all subsystems, (ii) system parameters being kept confidential and highly secure, 
%\cite{wang2020locational}, 
(iii) limitation on financial or instrument support, making it very challenging for attackers to access the target system's measurements. %\cite{li2022online}. 
Consequently, an attacker with limited system knowledge can generate stealthy attacks by min-cut problem \cite{hegazy2022online}.

%In practice, the system parameters are highly secured and kept confidential, which makes it hard for an attacker to have full knowledge \cite{wang2020locational}. Hence, with different advanced techniques, they try to collect information from the black-box to the grey-box level. Liu \emph{et. al.} \cite{liu2011false} showed that a minimum number of meters required for an attacker for successful stealthy attacks. Bi \emph{et. al.}\cite{bi2014using} proposed attack and defense simultaneously against the attacker with partial knowledge of the system. Another study of Suo \emph{et. al.} \cite{sou2011electric} addressed the possible way to find the minimum number of meters required to compromise the system.  

%Adversary knowledge is constituted by a set of information that can be exploited to increase the effectiveness of its attack. Such knowledge can be any type of information about the smart grid infrastructure, its components and operations. Adversary knowledge can be at different levels: (i) white-box: full knowledge about the underlying infrastructure, (ii) gray-box: limited knowledge about the underlying infrastructure, (iii) black-box: zero-knowledge about the underlying infrastructure . It is hard for attackers to have full knowledge about their target systems \cite{liu2011false}. Hence, with different advanced techniques, they try to collect information in order to reach from black-box to the grey-box level.

Table~\ref{tab:adversary-knowledge} shows the adversary knowledge levels and associated papers. We did not find any paper at black-box level with zero knowledge of the underlying system. Most of the reviewed articles considered the traditional bad data detection mechanism as their baseline for the generation of the attack vectors.

\begin{table}
\footnotesize
\centering
\caption{The knowledge of the adversary}
\begin{tabular}{|l|l|}
\hline
\textbf{Knowledge Level} & \textbf{Papers} \\ \hline \hline
        
\textbf{Full Knowledge} &  \begin{tabular}[c]{@{}l@{}} 
\cite{qiu2021cyber},
\cite{li2020deep}, 
\cite{li2022adaptive},
\cite{jiang2019location},
\cite{boyaci2022infinite}, \\
\cite{haghshenas2022temporal},
\cite{hegazy2022online},
\cite{li2022online},
\cite{fanlocational},
\cite{james2018online}
\end{tabular}\\ \hline

\textbf{Limited Knowledge} &  \begin{tabular}[c]{@{}l@{}} 
\cite{wang2020locational}, 
\cite{luo2018observer},
\cite{boyaci2021joint},
\cite{mukherjee2022deep}, 
\cite{vukovic2013detection}, 
\cite{luo2019detection},\\
\cite{khalafi2021intrusion}, 
\cite{shereen2022detection},
\cite{james2018online},
\cite{khalaf2018joint},
\cite{Hallaji2022Astream}
\end{tabular}\\ \hline

\end{tabular}
\label{tab:adversary-knowledge}
\end{table}

\subsection{Result Analysis}

Machine learning-based algorithms employ several key metrics to measure their performance. These evaluation metrics vary based on the problem under study and should be carefully selected in order to provide maximum insights. Some popular classification metrics include Accuracy, Precision, F1-Score, Recall, and Area Under the Curve (AUC), explained in the following:

\begin{itemize}
    \item[-] \textbf{Accuracy}: Accuracy is the most common parameter while evaluating performance. It is a fraction of the total number of correct predictions over the total number of samples. It is defined as: 
    \begin{equation}
        \label{eq:accuracy}
        Accuracy = \frac{TP + TN}{TP + TN + FP + FN}
    \end{equation}
    % \begin{center}
    %     \[Accuracy = \frac{TP + TN}{TP + TN + FP + FN} \]
    % \end{center}
    where TP stands for true positive, FP for false positive, TN for true negative, and FN for false negative. \\
    
    \item[-] \textbf{Precision}: Precision is used to measure a model's capacity to avoid FP. It is a fraction of TP and all positive predictions: 
    \begin{equation}
        \label{eq:Precision}
        Precision = \frac{TP}{TP+FP}
    \end{equation}
    % \begin{center}
    %     \[Precision = \frac{TP}{TP+FP} \]
    % \end{center}
    
    \item[-] \textbf{Recall}: 
     It is a fraction of TP measuring the percentage of true positives out of all actual positive instances: 
    
    \begin{equation}
        \label{eq:Recall}
        Recall = \frac{TP}{TP+FN}
    \end{equation}
    
    % \begin{center}
    %     \[Recall = \frac{TP}{TP+FN} \]
    % \end{center}
    \item[-] \textbf{F1-Score}: F1-score represents the balanced harmonic mean of precision and recall, serving as a metric for the model's overall performance. Its purpose is to ensure that precision and recall are considered equally:
    
    \begin{equation}
        \label{eq:f1score}
        F1_{score} = 2 \times \frac{Precision \times Recall}{Precision + Recall}
    \end{equation}
    
    % \begin{center}
    %     \[F1_{score} = 2 \times \frac{Precision \times Recall}{Precision + Recall} \]
    % \end{center}
    
    \item[-] \textbf{AUC}: AUC is a widely used evaluation metric in binary classification tasks that gauges the model's efficacy in discriminating between positive and negative instances:

    \begin{equation}
        \label{eq:AUC}
        AUC = \frac{S_{p}-n_{p}(n_{n}+1)/2}{n_{p} \times n_{n}}
    \end{equation}
    
    % \begin{center}
    %     \[AUC = \frac{S_{p}-n_{p}(n_{n}+1)/2}{n_{p} \times n_{n}} \]
    % \end{center}
    where $S_{p}$, $n_{n}$, and $n_{p}$ denote the sum of classified samples as positive, negative samples, and positive samples, respectively.
\end{itemize}

Our survey spans over several IEEE standard bus systems. We list the results of the selected studies based on various evaluation metrics and their utilized standard bus systems in Table~\ref{tab:localization_comparison}. As of using different datasets (not publicly available) and evaluation setups, we cannot perform a fair and complete comparison among all the selected works based on their results. However, there are cases adopting the same datasets, such as~\cite{wang2020locational},~\cite{fanlocational}, to conduct their evaluations. As mentioned previously, the main shortcoming of the most of these researches is the lack of evaluation within real environments or based on ground truth datasets. 

%We compared the selected studies based on various key metrics for evaluation summarized in Table~\ref{tab:localization_comparison}. This comparison put a constraint on comparing studies chosen due to the unavailability of the public dataset. Wang \emph{et. al.} \cite{wang2020locational} and Fan \emph{et. al.} \cite{fanlocational} used the same dataset for the evaluation of their studies. Due to this reason existing studies may fail to reproduce the results in a real scenario. This failure is also due to the simulation environment. There are some assumptions made from the perspective of an adversary this assumption can also be the source of failure in real systems.

%\input{table_test}

% \usepackage{multirow}
\begin{table*}[]
\centering
\caption{Performance summary in terms of evaluation Metrics and detection time}
\label{tab:localization_comparison}

\begin{tabular}{cclllllll}
\hline
\multirow{2}{*}{} & \multicolumn{1}{l}{\multirow{2}{*}{\textbf{}}} & \multirow{2}{*}{\textbf{Article}} & \multicolumn{6}{c}{\textbf{Metrics}} \\ \cline{4-9} 
 & \multicolumn{1}{l}{} &  & Accuracy & Precision & F1-Score & Recall & AUC & Detection Time \\ \hline
\multirow{25}{*}{\textbf{Localization}} & \multirow{10}{*}{\textbf{IEEE-14}} & Wang \emph{et. al.} \cite{wang2020locational} & 97.3 R & 99.66 & 99.75 & 99.84 & - & CNM \\
 &  & Mukherjee \emph{et. al.} \cite{mukherjee2022deep} & 94.01 R & 99.63 & 99.72 & 99.82 & - & $<$ $\mu$ sec \\
 &  & Lee \emph{et. al.} \cite{Lee2021} & - & 99.54 & 99.53 & 99.54 & Yes & - \\
 &  & Shereen \emph{et. al.} \cite{shereen2022detection} & CNM & - & - & - & Yes & $<$ milli sec \\
 &  & Jiang \emph{et. al.} \cite{jiang2019location} & 96.91* & - & - & - & - & - \\
 &  & Li \emph{et. al.} \cite{li2021locating} & 100 & - & - & - & - & - \\
 &  & Hegazi \emph{et. al.} \cite{hegazy2022online} & 95 R & 99.43 & 99.3 & 99.64 & - & - \\
 &  & Han-2023 & 98.3 & 99.8 & 92.6 & 86.4 & Yes & - \\
 &  & Li \emph{et. al.} \cite{li2022online} & CNM & CNM & CNM & CNM & - & - \\
 &  & Fan \emph{et. al.} \cite{fanlocational} & - & $>$99 & $>$99 & $>$99 & Yes & - \\ \cline{2-9} 
 & \multirow{2}{*}{\textbf{IEEE-57}} & Boyaci \emph{et. al.} \cite{boyaci2021joint} & - & - & 79.53 & - & - & 2.76 milli sec \\
 &  & Mohammadpourfard \emph{et. al.} \cite{mohammadpourfard2021attack} & - & 99 & 99 & 99 & - & - \\ \cline{2-9} 
 & \multirow{10}{*}{\textbf{IEEE-118}} & Wang \emph{et. al.} \cite{wang2020locational} & 93.3 R & 99.08 & 99.4 & 99.71 &  & 100 $\mu$ sec \\
 &  & Boyaci \emph{et. al.} \cite{boyaci2021joint} & - & - & 83 & - & - & 2.81 milli sec \\
 &  & Mukherjee \emph{et. al.} \cite{mukherjee2022deep} & 93.3 R & 99.8 & 99.7 & 99.61 & - & 200 $\mu$ sec \\
 &  & Vukovic \emph{et. al.} \cite{vukovic2013detection} & - & - & - & - & - & - \\
 &  & Li \emph{et. al.} \cite{li2022online} & CNM & CNM & CNM & CNM & - & - \\
 &  & Hasnat \emph{et. al.} \cite{hasnat2021} & $>$90 & - & - & - & - & - \\
 &  & Haghshenas \emph{et. al.} \cite{haghshenas2022temporal} & $>$95 & - & - & - & - & 0-20 time instance \\
 &  & Hegazi \emph{et. al.} \cite{hegazy2022online} & 93 R & 99.93 & 99.95 & 99.96 & - & - \\
 &  & Han \emph{et. al.} \cite{han2023false} & 96.4 & 98.9 & 83.3 & 72 & - &  \\
 &  & Fan \emph{et. al.} \cite{fanlocational} & - & $>$99 & $>$99 & $>$99 & Yes & - \\ \cline{2-9} 
 & \multirow{3}{*}{\textbf{IEEE-300}} & Han \emph{et. al.} \cite{han2023false} & 95.8 & 96.7 & 80.2 & 68.5 & - & - \\
 &  & Boyaci \emph{et. al.} \cite{boyaci2022infinite} & - & - & 93.33 & - & - & 2.94 milli sec \\
 &  & Boyaci \emph{et. al.} \cite{boyaci2021joint} & - & - & 79.03 & - & - & 2.94 milli sec \\ \hline
\end{tabular}
\vspace{1ex}
\\
{\raggedright CNM: Considered but not explicitly mentioned, *: Average, R: Row Accuracy or RACC, -: did not considered  \par}
\end{table*}

%gab: What is the meaning of AC and DC?

\subsection{AC/DC Power System}
%In alternating current (AC), the electrons flow bidirectionally and can be represented using a sine curve \cite{theraja2005textbook}. On the other hand, Direct Current (DC) has only one direction. 
While Alternating Current (AC) is suitable for distribution of power over longer distances, Direct Current (DC) is mostly used in batteries, electric vehicles, and low voltage power systems.
As discussed earlier, the main objective of most of the selected papers is state estimations in power systems \cite{wang2020locational}\cite{mukherjee2022deep}\cite{luo2019detection}. The complexities of a state estimation task depend on a variety of factors. One of the main factors is when the authors choose between AC or DC state estimations. DC estimation is simpler and gives an overview of the underlying system. It can be used in the preliminary phase of the study to capture the initial pictures. However, to truly replicate a power system, AC estimation is required. AC estimation is a non-linear problem and is a significantly complicated scenario compared to DC estimation. In terms of the sampling time, DC systems has lower sampling rate providing more time for attackers to compromise the system. Hence, deploying various hidden attack vectors against DC systems is considerably easier compared to AC systems that send faster snapshot of readings. Table~\ref{tab:ac-dc} categorizes the reviewed researches based on the type of their analyzed system. 

%==Muhammad's comment: 
%-> We can add the comparison between the DC System vs. AC in terms of the sampling time. The sampling rate in DC is lower. An attacker can inject an attack vector into the system without being detected before the following sample. AC sends a higher snapshot of readings which gives less time for an attacker to perform his attack. 

\begin{table}
\footnotesize
\centering
\caption{Categorization of the papers based on AC/DC architecture}
\begin{tabular}{|l|l|}
\hline
\textbf{System Type} & \textbf{Papers} \\ \hline\hline

\textbf{AC System} & \begin{tabular}[c]{@{}l@{}}
\cite{luo2018observer},
\cite{luo2019detection},
\cite{boyaci2021joint},
\cite{luo2020interval},
\cite{smith2009event},
\cite{wang2019detection},
\cite{smith2009event},\\
\cite{luo2019detection},
\cite{ganjkhani2021integrated},
\cite{hasnat2020detection},
\cite{qiu2021cyber},
\cite{wang2013detection},
\cite{li2020deep},
\cite{hasnat2022graph},\\
\cite{li2022adaptive},
\cite{boyaci2022infinite},
\cite{haghshenas2022temporal},
\cite{li2022online},
\cite{james2018online}
\end{tabular}\\ \hline
 
\textbf{DC System} & \begin{tabular}[c]{@{}l@{}}
\cite{wang2020locational}, 
\cite{mukherjee2022deep}, 
\cite{shi2018pdl},
\cite{mohammadpourfard2021attack},
\cite{jiang2019location},
\cite{shereen2022detection},
\cite{hasnat2021},\\
\cite{li2021locating},
\cite{gao2022fast},
\cite{hegazy2022online},
\cite{han2023false},
\cite{fanlocational},
\cite{sakhnini2021physical}
\end{tabular}\\ \hline

\end{tabular}
\label{tab:ac-dc}
\end{table} 

\subsection{Investigated FDI Attack Types}
A number of FDI attacks were studied in the reviewed studies to investigate the resilience of proposed approaches in the presence of an adversary. Each FDI attack has different features and profile. In the following, we describe the investigated FDI attacks. Table~\ref{tab:fdia-attack-types} lists the investigated FDI attacks and corresponding papers.

\subsubsection{Ramp Attack}
%Ramp attack enjoys a particular type of FDIA.
Ramp attacks cause power instability and equipment damages ~\cite{wang2020detection}. In these attacks, the attacker aims to evade the attack detection from linear-prediction-based models by introducing bad data to the smart grid data stream. Linear detection models fail to detect ramp attacks due to their gradual increase in nature. A successful ramp attack can cause significant financial and operational damages. 

%the successful attack results in financial losses for the utilities by fake-line congestion alarms and unprofitable dispatch \cite{wang2020detection}. 

\subsubsection{Replay Attack}
A replay attack, sometimes called a data replay attack, in which an adversary observes and records the data stream from smart grids meters and re-injects these measurements later to disrupt the system~\cite{hasnat2021detecting}.  
%Hasnat \emph{et. al.} \cite{hasnat2021detecting} mathematically described replay attack in Equation~\ref{eq:replay_attack}.

%\begin{equation}
%\label{eq:replay_attack}
%    x_{replay_i} = \begin{cases} 
%      x_{k}(t^{'} + t - t_{start}) + n_{i}(t) &  t_{start}\leq t \leq  t_{end} \\
%      x_{i}(t) + n_{i}(t) & otherwise
%   \end{cases}
%\end{equation}

%Where subscript $i$ belongs to the set of busses under attack, $k$ belongs to the set of busses for which an adversary has access to observe, $x$ time series of any electrical attribute such as voltage, Power, etc., $t^{'}$ starting of the recording time and $n_{i}(t)$ is Additive White Gaussian Noise.

\subsubsection{Covert Attack}
A covert attack is a type of FDI attack that can bypass the traditional BDD systems and remains unobservable for an extended time period to deteriorate the normal operation of the smart grids~\cite{li2020deep}.

\subsubsection{Denial of Service Attack}
Denial of Service (DoS) attacks cause smart grids services and measurements unavailability which lead to the unobservability of the underlying system. Overall, DoS attacks deteriorate the normal operation of smart grids. 

%DoS attack is the denial of service or unavailability of any measurement. When sensor/s data is unavailable, it can lead to the system's unobservability, which impacts the reliable operation of the system. 

%The denial of service attack is considered in \cite{hasnat2020detection}\cite{hasnat2021detecting}\cite{hasnat2021}. 

\subsubsection{Delay Attack}
A delay attack is carried out by the adversary to introduce non-synchronous behavior in the time-stamped measurements. Mathematically it can be represented as $c(t) = x(n_{A},t-t_{d})$ where $c(t)$ is corrupted measurement delayed by the amount of time $t_{d}$ and $n_{A}$ is the bus instance under attack. Time Synchronization Attack (TSA) is considered as a delay attack against synchronized PMU devices to change the phase angles of the measurements~\cite{shereen2022detection}.

\subsubsection{Random FDI Attack}
A random attack is an attack in which an attacker tries to find attack vectors leading the underlying smart grid controllers and operators to mis-estimate the state of the system~\cite{liu2011false}.  \\

%There could be different methodologies for FDI attacks, but the attacker's main objective in different circumstances is to achieve its malafide intentions. The summary of FDI attack types investigated by the researchers is presented in Table~\ref{tab:fdia-attack-types}.

% \usepackage{graphicx}
\begin{table}
\footnotesize
\centering
\caption{Types of FDI Attacks}
\label{tab:fdia-attack-types}
%\resizebox{\columnwidth}{!}{%
\begin{tabular}{|l|l|}
\hline
\textbf{FDI Attack Type} & \textbf{Papers} \\ \hline
\hline
Covert Attack& 
\cite{luo2018observer}\cite{luo2019detection}\cite{luo2020interval}\cite{li2022online}
 \\ \hline

Ramp Attack& 
\cite{wang2019detection}\cite{hasnat2021detecting}\cite{haghshenas2022temporal}\cite{wang2019online}
 \\ \hline
DoS Attack& 

\cite{hasnat2020detection}\cite{qiu2021cyber}
\cite{hasnat2021detecting}\cite{hasnat2022graph}\cite{hasnat2021} \\ \hline
Replay Attack& 
\cite{boyaci2021joint}\cite{ganjkhani2021integrated}\cite{hasnat2020detection}\cite{hasnat2022graph}\cite{hasnat2021detecting}\cite{hasnat2021}
\\ \hline
Delay Attack& \cite{hasnat2022graph}\cite{hasnat2021}\\ \hline
Time Synchronization Attack& \cite{shereen2022detection} \\ \hline
Random Attack & \cite{luo2019detection}  \\ \hline

\end{tabular}%
%}
\end{table}

%\subsection{Mitigation}
% summarize mitigation strategies extracted from our listed papers

%\subsubsection{Online vs. Offline Analysis}

\section{Open issues and research directions}
\label{sec:open-issues}

In this section, we describe current research challenges and possible future directions about the detection and localization of FDI attacks in smart grid. We believe that the following topics worth more investigation as future research directions in this field.

%\subsection{Data-Driven Modeling}

\begin{itemize}
%gab: this should be well justified. Of course, I agree that real measurements are better than simulations... but why? we need to identify the point maybe citing a few papers.
    \item[-] \textbf{Ground truth datasets.} Measurements from real world scenarios are extremely difficult to set-up due to privacy constraints and data confidentiality of critical infrastructures. While simulations provide a quick and easy to achieve test-bed for FDI cybersecurity attacks, they might not be the perfect representation of the system---this being true in particular for smart grid environments. 
    \item[-] \textbf{Real-world smart grid environments.} One of the best but expensive option is testing the proposed approach in a real-world environment, considering real constraints, noise and real FDI cyber-attacks. A real world deployment involving a real smart-grid can definitely improve the robustness of the findings along with enhancing its efficiency and performance.     
    \item[-] \textbf{Scalability.} Current approaches focus on specific, small scale, simulated scenarios. We highlight the importance of testing the proposed approaches in more complex and large-scale smart grids. This will improve robustness, scalability and comprehensiveness of FDI attack detection and localization approaches. 
    \item[-] \textbf{High-end FDI attacks:} FDI attacks might exploit recent advances on ML and AI, in particular, resorting to adversarial machine learning, i.e., generating attack patterns that are specifically designed to be undetected to AI-based detection algorithm.
    \item[-] \textbf{Real-time analysis.} Detecting, localizing and mitigating FDI attacks in real-time is becoming of great importance in order to reduce the adversary's advantage of performing malicious activities during the (stealthy) period of persistence of the adversary inside the smart grid ICT infrastructure. Hence, moving from offline to online detection and localization should be taken into account when designing new approaches. 
    \item[-] \textbf{Dataset quality improvement.} Dataset quality is of paramount importance to ensure model robustness and reliability. For data-driven models, training, evaluation and testing datasets need to be significantly well-elaborated. Dataset require to be up-to-date including new FDI attacks, while addressing imbalanced data and normalization, thus improving model quality (higher accuracy, precision and recall).     
    \item[-] \textbf{Differentiating cyber from physical.} Designing a model for detecting, localizing and distinguishing cyber from physical attacks is becoming more and more a priority due to the different countermeasures required for the defense~\cite{hasnat2021detecting}. For example, a PMU generating faulty measurements should be treated in a completely  way from a compromised PMU. For this purpose, a group of criteria should be defined and continuously updated to ensure more accurate discrimination. 
    \item[-] \textbf{Anomaly detection.} Many contributions from the literature dealing with FDI detection and localization are related to anomaly detection, i.e., the attack pattern is treated as an anomaly respect to a previously generated ground truth. Anomaly detection techniques focusing on 0-day (unknown) attacks might be considered to increase the detectability of 0-days in the domain of FDI attacks.
   \item[-] \textbf{Data correlation from various sources.} Correlating data from different sources, such as network data, topology information, and PMU characteristics, enhances FDI attack detection and localization to efficiently extract information for location detection of the attacks, thus increasing the overall performance of the mitigation activities. 
    \item[-] \textbf{False positive and false negative.} Reducing false positive and false negative can significantly improve the reliability and effectiveness of FDI attack detection and localization approaches in real-world environments. For data-driven models, this can be partially achieved by improving the training, evaluation and testing data quality. Data correlation also can ensure more precised and accurate detection and localization. 
    \item[-] \textbf{Graph signal processing (GSP) methods.} Potential improvements on GSP methods involve: (i) investigating the effect of various noises on their performance, (ii) applying GSP methods to dynamic graphs enabling them to detect and localize the group of FDI attacks causing changes in the network topology, (iii) using graph signal characteristics and vertex attributes as features in data-driven modelings, (iv) improving vertex-frequency energy distribution (VFED) algorithms to reduce computational complexities while enhancing detection and localization accuracy, (v) applying VFED algorithms to smal-size systems for stress localization. 
   \item[-] \textbf{Stress classification.} Comprehensive classification and characterisation of various stresses, and designing corresponding investigation and analysis mechanisms ensure more accurate and reliable prevention, prediction, detection, localization and mitigation. 
    \item[-] \textbf{Dynamic topology.} FDI attacks detection and localization in dynamic topology smart grids is another future direction to improve current approaches and keep their coherency with advancing technologies. This requires context awareness for the detection and localization algorithms to actively fetch underlying system features and characteristics to adapt with changing variables. Evaluating such approaches on multiple systems with varying topology and characteristics will significantly enhance its adoption within such environments. 
    \item[-] \textbf{Graphical network topology.} Integrating graphical network topology into the implemented version of current FDI attack detection and localization approaches can be another useful and valuable future direction enhancing their applicability and facilitating their analysis by specialists.
    % \item[-] \textbf{Other cyberattacks and faults:}
    % extending current FDI attack detection and localization approaches to analyze wider range of cyberattacks and physical faults worth more investigation.
%gab: this is to general...
    % \item[-] \textbf{Nonlinear analysis.} For mathematical modeling approaches, more analysis, investigation and optimization on nonlinear components of the proposed models can be an interesting future research direction. 
\end{itemize}

\section{Conclusion}
\label{sec:conclusion}
Smart grid state estimation integrity is critical for its proper operations. Adversaries adopt various advanced techniques to affect the (cyber) internal state, which in turn, reflects on the physical domain having a detrimental effect on both the security and safaty of the system. Among all, FDI attacks are the most powerful techniques for attackers to reach their objectives. Unfortunately, conventional detection methods (e.g., residual-based methods) are not able to detect such attacks due to recent advances on the attacker side. While detection of such attacks is still a significant challenge, minor effort have been devoted for localization---due to the complexity of the problem. We investigated the current state of the art on smart grids cyber-attacks, focusing on FDI attacks detection and localization. First, we provided a background on smart grids architecture and current cyber-security challenges. Next, after categorizing our list of selected papers into data-driven and mathematical modeling categories, we described their proposed approaches and employed techniques. Accordingly, we provided our key findings and analytical insights regarding the employed localization techniques, evaluation scenarios, obtained results, investigated FDI attack types, etc. Finally, we discussed about open issues in FDI attack joint detection and localization, and potential future research directions, while drawing some final remarks.

%In this work, we have investigated the current state of the art on smart grids cyber-attacks, focusing on FDI attacks, detection and localization. We have provided a background on smart grids architecture and cyber-security challenges. Next, we have categorized all the works considered in our analysis by the adopted methodology. Moreover, we have categorized the considered works as per evaluation scenarios, data sources, adopted techniques, and datasets. Finally, we have discussed open issues and potential research directions, while drawing some final remarks.

\section*{Acknowledgements}
%The authors would like to thank the anonymous reviewers for their comments, that helped improving the manuscript quality.
This publication was supported by NPRP grant NPRP12C-0814-190012 from the Qatar National Research Fund (a member of Qatar Foundation). The findings achieved herein are solely the responsibility of the authors. The authors also thank the Qatar National Library (QNL) for supporting this research.

\bibliographystyle{IEEEtran}
\balance
\bibliography{main}

\begin{thebibliography}{10}
\providecommand{\url}[1]{#1}
\csname url@samestyle\endcsname
\providecommand{\newblock}{\relax}
\providecommand{\bibinfo}[2]{#2}
\providecommand{\BIBentrySTDinterwordspacing}{\spaceskip=0pt\relax}
\providecommand{\BIBentryALTinterwordstretchfactor}{4}
\providecommand{\BIBentryALTinterwordspacing}{\spaceskip=\fontdimen2\font plus
\BIBentryALTinterwordstretchfactor\fontdimen3\font minus
  \fontdimen4\font\relax}
\providecommand{\BIBforeignlanguage}[2]{{%
\expandafter\ifx\csname l@#1\endcsname\relax
\typeout{** WARNING: IEEEtran.bst: No hyphenation pattern has been}%
\typeout{** loaded for the language `#1'. Using the pattern for}%
\typeout{** the default language instead.}%
\else
\language=\csname l@#1\endcsname
\fi
#2}}
\providecommand{\BIBdecl}{\relax}
\BIBdecl

\bibitem{yu2016smart}
X.~Yu and Y.~Xue, ``Smart grids: A cyber--physical systems perspective,''
  \emph{Proceedings of the IEEE}, vol. 104, no.~5, pp. 1058--1070, 2016.

\bibitem{pillitteri2014guidelines}
V.~Y. Pillitteri and T.~L. Brewer, ``Guidelines for smart grid cybersecurity,''
  2014.

\bibitem{musleh2019survey}
A.~S. Musleh, G.~Chen, and Z.~Y. Dong, ``A survey on the detection algorithms
  for false data injection attacks in smart grids,'' \emph{IEEE Transactions on
  Smart Grid}, vol.~11, no.~3, pp. 2218--2234, 2019.

\bibitem{deng2016false}
R.~Deng, G.~Xiao, R.~Lu, H.~Liang, and A.~V. Vasilakos, ``False data injection
  on state estimation in power systems—attacks, impacts, and defense: A
  survey,'' \emph{IEEE Transactions on Industrial Informatics}, vol.~13, no.~2,
  pp. 411--423, 2016.

\bibitem{aoufi2020survey}
S.~Aoufi, A.~Derhab, and M.~Guerroumi, ``Survey of false data injection in
  smart power grid: Attacks, countermeasures and challenges,'' \emph{Journal of
  Information Security and Applications}, vol.~54, p. 102518, 2020.

\bibitem{cui2020detecting}
L.~Cui, Y.~Qu, L.~Gao, G.~Xie, and S.~Yu, ``Detecting false data attacks using
  machine learning techniques in smart grid: A survey,'' \emph{Journal of
  Network and Computer Applications}, vol. 170, p. 102808, 2020.

\bibitem{guan2015comprehensive}
Z.~Guan, N.~Sun, Y.~Xu, and T.~Yang, ``A comprehensive survey of false data
  injection in smart grid,'' \emph{International Journal of Wireless and Mobile
  Computing}, vol.~8, no.~1, pp. 27--33, 2015.

\bibitem{husnoo2022false}
M.~A. Husnoo, A.~Anwar, N.~Hosseinzadeh, S.~N. Islam, A.~N. Mahmood, and
  R.~Doss, ``False data injection threats in active distribution systems: A
  comprehensive survey,'' \emph{Future Generation Computer Systems}, 2022.

\bibitem{liang2016review}
G.~Liang, J.~Zhao, F.~Luo, S.~R. Weller, and Z.~Y. Dong, ``A review of false
  data injection attacks against modern power systems,'' \emph{IEEE
  Transactions on Smart Grid}, vol.~8, no.~4, pp. 1630--1638, 2016.

\bibitem{reda2022comprehensive}
H.~T. Reda, A.~Anwar, and A.~Mahmood, ``Comprehensive survey and taxonomies of
  false data injection attacks in smart grids: Attack models, targets, and
  impacts,'' \emph{Renewable and Sustainable Energy Reviews}, vol. 163, p.
  112423, 2022.

\bibitem{sayghe2020survey}
A.~Sayghe, Y.~Hu, I.~Zografopoulos, X.~Liu, R.~G. Dutta, Y.~Jin, and
  C.~Konstantinou, ``Survey of machine learning methods for detecting false
  data injection attacks in power systems,'' \emph{IET Smart Grid}, vol.~3,
  no.~5, pp. 581--595, 2020.

\bibitem{AOUFI2020}
\BIBentryALTinterwordspacing
S.~Aoufi, A.~Derhab, and M.~Guerroumi, ``Survey of false data injection in
  smart power grid: Attacks, countermeasures and challenges,'' \emph{Journal of
  Information Security and Applications}, vol.~54, p. 102518, 2020. [Online].
  Available:
  \url{https://www.sciencedirect.com/science/article/pii/S2214212619310713}
\BIBentrySTDinterwordspacing

\bibitem{youssef2018false}
E.-N.~S. Youssef and F.~Labeau, ``False data injection attacks against state
  estimation in smart grids: Challenges and opportunities,'' in \emph{2018 IEEE
  Canadian Conference on Electrical \& Computer Engineering (CCECE)}.\hskip 1em
  plus 0.5em minus 0.4em\relax IEEE, 2018, pp. 1--5.

\bibitem{gopstein2021nist}
A.~Gopstein, C.~Nguyen, C.~O'Fallon, N.~Hastings, D.~Wollman \emph{et~al.},
  \emph{NIST framework and roadmap for smart grid interoperability standards,
  release 4.0}.\hskip 1em plus 0.5em minus 0.4em\relax Department of Commerce.
  National Institute of Standards and Technology, 2021.

\bibitem{Anuradha2013IEEEVision}
A.~M. Annaswamy and M.~Amin, ``Ieee vision for smart grid controls: 2030 and
  beyond,'' \emph{IEEE Vision for Smart Grid Controls: 2030 and Beyond}, pp.
  1--168, 2013.

\bibitem{cardenas2019cyber}
A.~Cardenas and S.~Cruz, ``Cyber-physical systems security knowledge area,''
  \emph{The Cyber Security Body Of Knowledge (cybok)}, 2019.

\bibitem{konstantinou2016case}
C.~Konstantinou and M.~Maniatakos, ``A case study on implementing false data
  injection attacks against nonlinear state estimation,'' in \emph{Proceedings
  of the 2nd ACM workshop on cyber-physical systems security and privacy},
  2016, pp. 81--92.

\bibitem{parizad2021laboratory}
A.~Parizad and C.~Hatziadoniu, ``A laboratory set-up for cyber attacks
  simulation using protocol analyzer and rtu hardware applying semi-supervised
  detection algorithm,'' in \emph{2021 IEEE Texas Power and Energy Conference
  (TPEC)}.\hskip 1em plus 0.5em minus 0.4em\relax IEEE, 2021, pp. 1--6.

\bibitem{shereen2022detection}
E.~Shereen, R.~Ramakrishna, and G.~D{\'a}n, ``Detection and localization of pmu
  time synchronization attacks via graph signal processing,'' \emph{IEEE
  Transactions on Smart Grid}, 2022.

\bibitem{khalaf2018joint}
M.~Khalaf, A.~Youssef, and E.~El-Saadany, ``Joint detection and mitigation of
  false data injection attacks in agc systems,'' \emph{IEEE Transactions on
  Smart Grid}, vol.~10, no.~5, pp. 4985--4995, 2018.

\bibitem{ganjkhani2019novel}
M.~Ganjkhani, S.~N. Fallah, S.~Badakhshan, S.~Shamshirband, and K.-w. Chau, ``A
  novel detection algorithm to identify false data injection attacks on power
  system state estimation,'' \emph{Energies}, vol.~12, no.~11, p. 2209, 2019.

\bibitem{an2022data}
D.~An, F.~Zhang, Q.~Yang, and C.~Zhang, ``Data integrity attack in dynamic
  state estimation of smart grid: Attack model and countermeasures,''
  \emph{IEEE Transactions on Automation Science and Engineering}, vol.~19,
  no.~3, pp. 1631--1644, 2022.

\bibitem{ibrahem2020pmbfe}
M.~I. Ibrahem, M.~M. Badr, M.~M. Fouda, M.~Mahmoud, W.~Alasmary, and Z.~M.
  Fadlullah, ``Pmbfe: Efficient and privacy-preserving monitoring and billing
  using functional encryption for ami networks,'' in \emph{2020 international
  symposium on networks, computers and communications (ISNCC)}.\hskip 1em plus
  0.5em minus 0.4em\relax IEEE, 2020, pp. 1--7.

\bibitem{alsharif2019epda}
A.~Alsharif, M.~Nabil, M.~M. Mahmoud, and M.~Abdallah, ``Epda: Efficient and
  privacy-preserving data collection and access control scheme for
  multi-recipient ami networks,'' \emph{IEEE Access}, vol.~7, pp.
  27\,829--27\,845, 2019.

\bibitem{na2021fake}
L.~Na, X.~Xiaohui, M.~Xiaoqin, M.~Xiangfu, and Y.~Peisen, ``Fake data injection
  attack detection in ami system using a hybrid method,'' in \emph{2021 IEEE
  Sustainable Power and Energy Conference (iSPEC)}.\hskip 1em plus 0.5em minus
  0.4em\relax IEEE, 2021, pp. 2371--2376.

\bibitem{Saxena2017secure}
N.~Saxena, B.~J. Choi, and S.~Grijalva, ``Secure and privacy-preserving
  concentration of metering data in ami networks,'' in \emph{2017 IEEE
  International Conference on Communications (ICC)}, 2017, pp. 1--7.

\bibitem{gholami2019cyber}
A.~Gholami, M.~Mousavi, A.~K. Srivastava, and A.~Mehrizi-Sani, ``Cyber-physical
  vulnerability and security analysis of power grid with hvdc line,'' in
  \emph{2019 North American Power Symposium (NAPS)}.\hskip 1em plus 0.5em minus
  0.4em\relax IEEE, 2019, pp. 1--6.

\bibitem{abur2004power}
A.~Abur and A.~G. Exposito, \emph{Power system state estimation: theory and
  implementation}.\hskip 1em plus 0.5em minus 0.4em\relax CRC press, 2004.

\bibitem{yuan2011modeling}
Y.~Yuan, Z.~Li, and K.~Ren, ``Modeling load redistribution attacks in power
  systems,'' \emph{IEEE Transactions on Smart Grid}, vol.~2, no.~2, pp.
  382--390, 2011.

\bibitem{davis2012power}
K.~R. Davis, K.~L. Morrow, R.~Bobba, and E.~Heine, ``Power flow cyber attacks
  and perturbation-based defense,'' in \emph{2012 IEEE third international
  conference on smart grid communications (SmartGridComm)}.\hskip 1em plus
  0.5em minus 0.4em\relax IEEE, 2012, pp. 342--347.

\bibitem{liu2011false}
Y.~Liu, P.~Ning, and M.~K. Reiter, ``False data injection attacks against state
  estimation in electric power grids,'' \emph{ACM Transactions on Information
  and System Security (TISSEC)}, vol.~14, no.~1, pp. 1--33, 2011.

\bibitem{wang2020locational}
S.~Wang, S.~Bi, and Y.-J.~A. Zhang, ``Locational detection of the false data
  injection attack in a smart grid: A multilabel classification approach,''
  \emph{IEEE Internet of Things Journal}, vol.~7, no.~9, pp. 8218--8227, 2020.

\bibitem{luo2018observer}
X.~Luo, Q.~Yao, X.~Wang, and X.~Guan, ``Observer-based cyber attack detection
  and isolation in smart grids,'' \emph{International Journal of Electrical
  Power \& Energy Systems}, vol. 101, pp. 127--138, 2018.

\bibitem{wang2020detection}
X.~Wang, X.~Luo, M.~Zhang, Z.~Jiang, and X.~Guan, ``Detection and isolation of
  false data injection attacks in smart grid via unknown input interval
  observer,'' \emph{IEEE Internet of Things Journal}, vol.~7, no.~4, pp.
  3214--3229, 2020.

\bibitem{boyaci2021joint}
O.~Boyaci, M.~R. Narimani, K.~R. Davis, M.~Ismail, T.~J. Overbye, and
  E.~Serpedin, ``Joint detection and localization of stealth false data
  injection attacks in smart grids using graph neural networks,'' \emph{IEEE
  Transactions on Smart Grid}, vol.~13, no.~1, pp. 807--819, 2021.

\bibitem{nudell2015real}
T.~R. Nudell, S.~Nabavi, and A.~Chakrabortty, ``A real-time attack localization
  algorithm for large power system networks using graph-theoretic techniques,''
  \emph{IEEE Transactions on Smart Grid}, vol.~6, no.~5, pp. 2551--2559, 2015.

\bibitem{mukherjee2022deep}
D.~Mukherjee, S.~Chakraborty, and S.~Ghosh, ``Deep learning-based multilabel
  classification for locational detection of false data injection attack in
  smart grids,'' \emph{Electrical Engineering}, vol. 104, no.~1, pp. 259--282,
  2022.

\bibitem{luo2020interval}
X.~Luo, Y.~Li, X.~Wang, and X.~Guan, ``Interval observer-based detection and
  localization against false data injection attack in smart grids,'' \emph{IEEE
  Internet of Things Journal}, vol.~8, no.~2, pp. 657--671, 2020.

\bibitem{siamak2020dynamic}
S.~Siamak, M.~Dehghani, and M.~Mohammadi, ``Dynamic gps spoofing attack
  detection, localization, and measurement correction exploiting pmu and
  scada,'' \emph{IEEE Systems Journal}, vol.~15, no.~2, pp. 2531--2540, 2020.

\bibitem{vukovic2013detection}
O.~Vukovi{\'c} and G.~D{\'a}n, ``Detection and localization of targeted attacks
  on fully distributed power system state estimation,'' in \emph{2013 IEEE
  International Conference on Smart Grid Communications (SmartGridComm)}.\hskip
  1em plus 0.5em minus 0.4em\relax IEEE, 2013, pp. 390--395.

\bibitem{luo2019detection}
X.~Luo, X.~Wang, X.~Pan, and X.~Guan, ``Detection and isolation of false data
  injection attack for smart grids via unknown input observers,'' \emph{IET
  Generation, Transmission \& Distribution}, vol.~13, no.~8, pp. 1277--1286,
  2019.

\bibitem{smith2009event}
M.~J. Smith and K.~Wedeward, ``Event detection and location in electric power
  systems using constrained optimization,'' in \emph{2009 IEEE Power \& Energy
  Society General Meeting}.\hskip 1em plus 0.5em minus 0.4em\relax IEEE, 2009,
  pp. 1--6.

\bibitem{ganjkhani2021integrated}
M.~Ganjkhani, M.~Gilanifar, J.~Giraldo, and M.~Parvania, ``Integrated cyber and
  physical anomaly location and classification in power distribution systems,''
  \emph{IEEE Transactions on Industrial Informatics}, vol.~17, no.~10, pp.
  7040--7049, 2021.

\bibitem{hasnat2020detection}
M.~A. Hasnat and M.~Rahnamay-Naeini, ``Detection and locating cyber and
  physical stresses in smart grids using graph signal processing,'' \emph{arXiv
  preprint arXiv:2006.06095}, 2020.

\bibitem{sakhnini2021physical}
J.~Sakhnini, H.~Karimipour, A.~Dehghantanha, and R.~M. Parizi, ``Physical layer
  attack identification and localization in cyber--physical grid: An ensemble
  deep learning based approach,'' \emph{Physical Communication}, vol.~47, p.
  101394, 2021.

\bibitem{shi2018pdl}
W.~Shi, Y.~Wang, Q.~Jin, and J.~Ma, ``Pdl: an efficient prediction-based false
  data injection attack detection and location in smart grid,'' in \emph{2018
  IEEE 42nd Annual Computer Software and Applications Conference (COMPSAC)},
  vol.~2.\hskip 1em plus 0.5em minus 0.4em\relax IEEE, 2018, pp. 676--681.

\bibitem{mohammadpourfard2021attack}
M.~Mohammadpourfard, I.~Genc, S.~Lakshminarayana, and C.~Konstantinou, ``Attack
  detection and localization in smart grid with image-based deep learning,'' in
  \emph{2021 IEEE international conference on communications, control, and
  computing technologies for smart grids (SmartGridComm)}.\hskip 1em plus 0.5em
  minus 0.4em\relax IEEE, 2021, pp. 121--126.

\bibitem{li2020deep}
D.~Li, P.~Ramanan, N.~Gebraeel, and K.~Paynabar, ``Deep learning based covert
  attack identification for industrial control systems,'' in \emph{2020 19th
  IEEE International Conference on Machine Learning and Applications
  (ICMLA)}.\hskip 1em plus 0.5em minus 0.4em\relax IEEE, 2020, pp. 438--445.

\bibitem{drayer2019cyber}
E.~Drayer and T.~Routtenberg, ``Cyber attack localization in smart grids by
  graph modulation (brief announcement),'' in \emph{International Symposium on
  Cyber Security Cryptography and Machine Learning}.\hskip 1em plus 0.5em minus
  0.4em\relax Springer, 2019, pp. 97--100.

\bibitem{khalafi2021intrusion}
Z.~S. Khalafi, M.~Dehghani, A.~Khalili, A.~Sami, N.~Vafamand, and
  T.~Dragi{\v{c}}evi{\'c}, ``Intrusion detection, measurement correction, and
  attack localization of pmu networks,'' \emph{IEEE Transactions on Industrial
  Electronics}, vol.~69, no.~5, pp. 4697--4706, 2021.

\bibitem{wang2013detection}
W.~Wang, L.~He, P.~Markham, H.~Qi, and Y.~Liu, ``Detection, recognition, and
  localization of multiple attacks through event unmixing,'' in \emph{2013 IEEE
  International Conference on Smart Grid Communications (SmartGridComm)}.\hskip
  1em plus 0.5em minus 0.4em\relax IEEE, 2013, pp. 73--78.

\bibitem{hasnat2021detecting}
M.~A. Hasnat and M.~Rahnamay-Naeini, ``Detecting and locating cyber and
  physical stresses in smart grids using the k-nearest neighbour analysis of
  instantaneous correlation of states,'' \emph{IET Smart Grid}, vol.~4, no.~3,
  pp. 307--320, 2021.

\bibitem{jiang2019location}
J.~Jiang, J.~Wu, C.~Long, and S.~Li, ``Location of false data injection attacks
  in power system,'' in \emph{2019 Chinese Control Conference (CCC)}.\hskip 1em
  plus 0.5em minus 0.4em\relax IEEE, 2019, pp. 7473--7478.

\bibitem{li2021locating}
B.~Li, Q.~Du, J.~Song, A.~Li, and X.~Ma, ``Locating false data injection
  attacks on smart grids using d-facts devices,'' in \emph{International
  Conference on Service-Oriented Computing}.\hskip 1em plus 0.5em minus
  0.4em\relax Springer, 2021, pp. 287--301.

\bibitem{hasnat2022graph}
M.~A. Hasnat and M.~Rahnamay-Naeini, ``A graph signal processing framework for
  detecting and locating cyber and physical stresses in smart grids,''
  \emph{IEEE Transactions on Smart Grid}, 2022.

\bibitem{qiu2021cyber}
Q.~Qiu, F.~Yang, and Y.~Zhu, ``Cyber-attack localisation and tolerant control
  for microgrid energy management system based on set-membership estimation,''
  \emph{International Journal of Systems Science}, vol.~52, no.~6, pp.
  1206--1222, 2021.

\bibitem{li2022adaptive}
Q.~Li, J.~Zhang, J.~Zhao, J.~Ye, W.~Song, and F.~Li, ``Adaptive hierarchical
  cyber attack detection and localization in active distribution systems,''
  \emph{IEEE Transactions on Smart Grid}, vol.~13, no.~3, pp. 2369--2380, 2022.

\bibitem{hasnat2021}
M.~R.-N. Md~Abul~Hasnat, ``Characterization and classification of cyber attacks
  in smart grids using local smoothness of graph signals,'' vol.~2, pp. 1--6,
  2021.

\bibitem{boyaci2022infinite}
O.~Boyaci, M.~R. Narimani, K.~Davis, and E.~Serpedin, ``Infinite impulse
  response graph neural networks for cyberattack localization in smart grids,''
  \emph{arXiv preprint arXiv:2206.12527}, 2022.

\bibitem{gao2022fast}
X.~Gao, X.~Yang, L.~Meng, and S.~Wang, ``Fast economic dispatch with false data
  injection attack in electricity-gas cyber--physical system: A data-driven
  approach,'' \emph{ISA transactions}, 2022.

\bibitem{hegazy2022online}
H.~I. Hegazy, A.~S.~T. Eldien, M.~M. Tantawy, M.~M. Fouda, and H.~A. TagElDien,
  ``Online location-based detection of false data injection attacks in smart
  grid using deep learning,'' in \emph{2022 IEEE International Conference on
  Internet of Things and Intelligence Systems (IoTaIS)}.\hskip 1em plus 0.5em
  minus 0.4em\relax IEEE, 2022, pp. 153--159.

\bibitem{han2023false}
Y.~Han, H.~Feng, K.~Li, and Q.~Zhao, ``False data injection attacks detection
  with modified temporal multi-graph convolutional network in smart grids,''
  \emph{Computers \& Security}, vol. 124, p. 103016, 2023.

\bibitem{li2022online}
D.~Li, N.~Z. Gebraeel, K.~Paynabar, and A.~S. Meliopoulos, ``An online approach
  to covert attack detection and indentification in power systems,'' \emph{IEEE
  Transactions on Power Systems}, 2022.

\bibitem{fanlocational}
X.~Fan, M.~Zhang, H.~Zeng, and C.~Shen, ``Locational detection of data
  integrity attacks with multi-gate mixture-of-experts in smart grid.''

\bibitem{haghshenas2022temporal}
S.~H. Haghshenas, M.~A. Hasnat, and M.~Naeini, ``A temporal graph neural
  network for cyber attack detection and localization in smart grids,''
  \emph{arXiv preprint arXiv:2212.03390}, 2022.

\bibitem{yong2016unified}
S.~Z. Yong, M.~Zhu, and E.~Frazzoli, ``A unified filter for simultaneous input
  and state estimation of linear discrete-time stochastic systems,''
  \emph{Automatica}, vol.~63, pp. 321--329, 2016.

\bibitem{Hallaji2022Astream}
E.~Hallaji, R.~Razavi-Far, M.~Wang, M.~Saif, and B.~Fardanesh, ``A stream
  learning approach for real-time identification of false data injection
  attacks in cyber-physical power systems,'' \emph{IEEE Transactions on
  Information Forensics and Security}, vol.~17, pp. 3934--3945, 2022.

\bibitem{james2018online}
J.~James, Y.~Hou, and V.~O. Li, ``Online false data injection attack detection
  with wavelet transform and deep neural networks,'' \emph{IEEE Transactions on
  Industrial Informatics}, vol.~14, no.~7, pp. 3271--3280, 2018.

\bibitem{jevtic2018physics}
A.~Jevtic, F.~Zhang, Q.~Li, and M.~Ilic, ``Physics-and learning-based detection
  and localization of false data injections in automatic generation control,''
  \emph{IFAC-PapersOnLine}, vol.~51, no.~28, pp. 702--707, 2018.

\bibitem{gu2013bad}
Y.~Gu, T.~Liu, D.~Wang, X.~Guan, and Z.~Xu, ``Bad data detection method for
  smart grids based on distributed state estimation,'' in \emph{2013 IEEE
  International Conference on Communications (ICC)}.\hskip 1em plus 0.5em minus
  0.4em\relax IEEE, 2013, pp. 4483--4487.

\bibitem{zonouz2012scpse}
S.~Zonouz, K.~M. Rogers, R.~Berthier, R.~B. Bobba, W.~H. Sanders, and T.~J.
  Overbye, ``Scpse: Security-oriented cyber-physical state estimation for power
  grid critical infrastructures,'' \emph{IEEE Transactions on Smart Grid},
  vol.~3, no.~4, pp. 1790--1799, 2012.

\bibitem{wang2019detection}
X.~Wang, X.~Luo, Y.~Zhang, and X.~Guan, ``Detection and isolation of false data
  injection attacks in smart grids via nonlinear interval observer,''
  \emph{IEEE Internet of Things Journal}, vol.~6, no.~4, pp. 6498--6512, 2019.

\bibitem{tierney2010new}
S.~F. Tierney, ``The new york independent system operator: A ten-year review,''
  \emph{New York ISO White Paper. Boston, MA}, 2010.

\bibitem{Lee2021}
B.~Li, Q.~Du, J.~Song, A.~Li, and X.~Ma, ``Locating false data injection
  attacks on smart grids using d-facts devices,'' in \emph{Service-Oriented
  Computing: 19th International Conference, ICSOC 2021, Virtual Event, November
  22–25, 2021, Proceedings}.\hskip 1em plus 0.5em minus 0.4em\relax Berlin,
  Heidelberg: Springer-Verlag, 2021, p. 287–301.

\bibitem{Lofberg2004}
J.~L{\"{o}}fberg, ``Yalmip : A toolbox for modeling and optimization in
  matlab,'' in \emph{In Proceedings of the CACSD Conference}, Taipei, Taiwan,
  2004.

\bibitem{wang2019online}
X.~Wang, D.~Shi, J.~Wang, Z.~Yu, and Z.~Wang, ``Online identification and data
  recovery for pmu data manipulation attack,'' \emph{IEEE Transactions on Smart
  Grid}, vol.~10, no.~6, pp. 5889--5898, 2019.

\end{thebibliography}
\end{document}